\documentclass{aa}
\usepackage{graphicx}
\usepackage{psfig}
\begin{document}
   \title{UV and FIR selected samples
of galaxies in the local Universe}
\subtitle{Dust extinction and star formation rates}
   \author{
J. Iglesias-P\'{a}ramo, V. Buat, J. Donas, A. Boselli, B. Milliard}

   \institute{Laboratoire d'Astrophysique de Marseille, BP8, 13376 Marseille 
cedex 12, France\\
\email{jorge.iglesias,veronique.buat,jose.donas,alessandro.boselli,bruno.milliard@oamp.fr}
}

   \date{Received ...; accepted ...}

   \abstract{
We have built two samples of galaxies selected at 0.2$\mu$m (hereafter UV)  
and 60 $\mu$m (hereafter FIR) covering a sky area of 35.36
deg$^2$. The UV selected sample contains 25 galaxies brighter than
$AB_{0.2}=17$. All of them, but one elliptical, 
are detected at 60 $\mu$m with a flux density larger or equal to  0.2~Jy. 
The UV counts are significantly lower than the euclidean extrapolation towards brighter fluxes of previous determinations.
The FIR selected sample contains 
42 galaxies brighter than $f_{60}$=0.6 Jy. Excepting four galaxies, all  
of them have a UV counterpart at the limiting magnitude 
$AB_{0.2}=20.3$~mag. The mean 
extinction derived from the analysis of the FIR
to UV flux ratio is $\sim 1$~mag for the UV selected sample 
and $\sim 2$~mag for the FIR selected one. 
For each sample we compare several indicators of the recent star formation 
rate (SFR) based on the FIR and/or the UV emissions and we find
linear relationships with slopes close to unity, meaning that no trend with the SFR exists when converting between each other. Various absolute calibrations for both samples are discussed in this paper.
A positive correlation between extinction and SFR is found when  
both samples are considered together although with a considerable scatter.
A similar result is obtained when using the SFR normalized to the optical surface of the galaxies.
}
  \titlerunning{Extinction and SFR in FIR and UV selected samples}

   \maketitle
%

\section{Introduction}

Tracing the star formation activity in galaxies at all redshifts is a
fundamental step towards the understanding of the formation and evolution of the Universe. The
star formation activity is commonly quantified by the Star Formation Rate (SFR)
defined as the stellar mass formed per unit of time. In order to 
efficiently constrain the models of galaxy  formation and evolution it is important to
measure
a  SFR as current as possible in order not to integrate on too large lookback
times. Quite naturally, the light emitted by young stars can be used to
measure this SFR. The most commonly used tracers of the SFR in galaxies, at least in the nearby Universe, are the UV, FIR and
H$\alpha$ emissions (e.g. Kennicutt 1998). In the present study we focus our attention on the UV and FIR emissions.

Concerning the FIR window, the IRAS mission provided us with several now well studied samples such as the Point Source Catalog (PSC, Joint IRAS
Science 1994), the Faint Source
Catalog (FSC, Moshir et al. 1990) and the redshift survey of the Point
Source Catalog (PSCz, Saunders et al. 2000).
The UV window,
less explored mostly because limited observations have been available until 
now (e.g. Donas et al. 1987; Deharveng et al. 1994; Milliard et al. 1992; Kinney et 
al. 1993; Bell \& Kennicutt 2001), has been extensively studied (e.g. Milliard et al. 1992; Donas 
et al. 1995; Treyer et al. 1998). This
situation should change dramatically in the near future once the GALEX data will be available.

The first step towards the determination of statistical properties of UV and FIR selected samples of galaxies is the determination of the galaxy counts and the luminosity function (LF). Accurate
determinations of these observables are important for constraining models of galaxy formation and evolution.
Nevertheless very few models predict the 
luminosity distribution and spectral energy distribution of galaxies over a 
large range of wavelength from UV to FIR (Totani \& Takeuchi 2002 and references therein; Xu et al. 1998). This is due, at least 
in part, to our poor knowledge of the dust extinction in the universe (amount of 
dust, dust emission, mechanism of stellar absorption on large scales). 
The reconstruction of the whole SED of galaxies from UV to FIR is difficult even in the well sampled nearby Universe because of the lack of multiwavelength data
on large and 
homogeneous samples of galaxies (Boselli et al. 2003; Flores et al. 1999; 
Rigopoulou et al. 2000; Cardiel et al. 2003). 
This situation should evolve  dramatically in 
the future with the planned observations of large fields with telescopes 
working at very different wavelengths (SIRTF, GALEX, VIMOS, ASTRO-F). 

The FIR to UV flux ratio has been proved to be the best indicator of the dust extinction in normal galaxies (e.g. Buat \& Xu 1996; Meurer et al. 1999), becoming a fundamental parameter in the
determination of their present SFR (e.g. Buat et al. 1999,2002; Hirashita et
al. 2003, hereafter HBI).

The aim of the present paper is to study the statistical properties of two samples extracted from the same area of the sky, selected according to UV and FIR criteria.
The two selections (UV and FIR) appear very complementary since 
the sample of UV selected galaxies will be
biased towards active star forming galaxies with low extinction,
and on  the contrary a FIR selection is likely to favor galaxies
with high extinction. 
An accurate knowledge of the statistical properties of UV and FIR selected samples is crucial for the analysis of similarly selected samples at higher $z$, where the lack of complementary data
prevents a precise determinations of the dust extinction and SFR.

With these two samples in hand, we
study the extinction and SFR related properties and their
dependence on the selection method. More precisely we address the following 
issues: (1) determine the mean extinction
of purely UV or FIR selected samples of galaxies, (2) 
observationally constrain models of galaxy formation and evolution, and (3) 
provide the best recipes for the determination of the SFR, valid for UV and/or FIR selected samples of galaxies.

The paper is organized as follows: Section~2 presents the FIR and UV
data and the region of the
sky where the data were taken. Sections~3 and 4 describe the
selection procedure of the UV and FIR selected samples. Section~5
discusses  the extinction properties of the samples and Section~6
is devoted to the comparison between the different SFR tracers. 
A final summary of the main results of the paper is
presented in Section~7.

\section {The FIR and UV data}

An accurate comparison between the properties of UV and FIR
galaxies can be done if both samples are extracted from the same
region of the sky. With this purpose we selected some fields covered by the
FOCA experiment (Milliard et al. 1991) at UV wavelengths (0.2$\mu$m). 
A total of 9 FOCA fields were chosen. The area of each field is circular with
a radius of 1.13~$\deg$. Accounting for some overlapping of the fields, a total 
area of 35.36~$\deg^{2}$ was covered. Table~\ref{tabla} shows the
basic properties of the selected fields. Four out of the nine observed fields are centered on nearby clusters of galaxies: m010 (Cancer), m028 (Coma), m067 (Abell~1367) and m050 (Virgo), and two of them
point to the external parts of the Coma cluster (m030, m031). The Cancer cluster
is however in Hubble flow (Gavazzi et al. 1991), thus its members are similar to field galaxies. The fractions of cluster galaxies in our UV selected and FIR selected (hereafter UVsel and FIRsel
respectively) samples are 64\%
and 50\% respectively. The possible contribution of the cluster environment to the properties of the UVsel and FIRsel galaxies are discussed in the text.

The detection and flux extraction of the UV objects in the FOCA plates
was carried out in an automatic way. Only UV sources with surface brightness (averaged over $15 \times 15$~arcsec$^{2}$) brighter than 2.8 times the sky $\sigma$ (the detection limit established by the automatic detection algorithm for each
frame, given in Col.~5 of Table~1)
were considered as detections. The automatic determination of the UV flux is accurate for point like sources but uncertain for extended objects, so the UV fluxes of the
FIRsel galaxies were determined in two ways: (1) for the galaxies which were not resolved by the automatic detection algorithm (i.e. only one UV source was detected) we used the UV flux provided by
the detection algorithm\footnote{The automatic algorithm performs aperture photometry around the selected UV sources using increasing radii. The adopted aperture is selected when the magnitude
stabilizes. More details about this procedure are
given in Moulinec (1989).}. (2) for the galaxies which were splitted in several sources (i.e. several H{\sc ii} regions were detected as individual sources) we performed aperture photometry by
hand. The optical images were used to set the size of the apertures in order to be sure that all the UV flux corresponding to the galaxies, even the most external star forming regions, was included in
the total UV flux. In both cases the apertures used to measure the UV fluxes are close to the optical diameters of the galaxies.
The average zero point uncertainty of the FOCA data
is about 0.2~mag. The spatial resolution of the FOCA frames is $\approx 20$~arcsecs FWHM and the $1\sigma$ astrometric uncertainty is $\approx 2.5$~arcsecs.

Concerning the FIR data, the IRAS all sky survey
ensures a total coverage of our selected FOCA fields. The total
photometric uncertainty of the FIR data of the IRAS mission is
$\approx 10$\% at 60 and 100$\mu$m. The positional uncertainty is variable and of elliptical shape with major axis $\approx 40$~arcsecs. The spatial resolution shows irregular shape and depends
on the bandwidth and coordinates of the object. In the direction where it is maximal being $\approx 4.5$~arcmin at 60$\mu$m (see Moshir et al. 1990 for details).

\section{The UV selected sample}

\subsection{Selection of UV galaxies}

The UVsel sample includes all galaxies with $AB_{0.2} < 17$~mag
\footnote{The UV magnitudes will always
be referred in the $AB$ system defined as $AB_{\nu} = -2.5
\log f_{\nu} - 48.6$, where $f_{\nu}$ is expressed in
erg~cm$^{-2}$~s$^{-1}$~Hz$^{-1}$.}. 
This limit is chosen in order to
have a good chance to find FIR counterparts ($f_{60} > 0.2$Jy for IRAS FSC). 
The combination of the adopted limits of both the UV and FIR flux densities sets the completeness limit for the extinction of our sample to $A_{0.2} = 0.7$~mag (see Eq.~1 in Sec.~5.2).

The determination of the UVsel sample is difficult
because of the high contribution of field stars to the total UV counts
extracted from the FOCA frames and the poor spatial resolution of the UV data 
(see below).  This might result in misclassification
of stars as galaxies. To get rid of this technical problem we followed this procedure:
\begin{enumerate}
\item At first we crosscorrelated
the sample of UV sources brighter than $AB_{0.2} = 17$~mag with
the NED database and obtained 25 galaxies. 
Fig~\ref{bonafide} shows the UV (from
the FOCA frames) and optical (from the DSS blue plates) images
of these galaxies. As can be seen from the figure, all
these galaxies show a clear extended structure and by no means can be misclassified as point like objects.\\

\item The remaining UV sources brighter than $AB_{0.2} = 17$~mag, a total of
375, were not found to be classified as galaxies in the NED database nor in the 2MASS extended source catalog. These UV
sources do not show an extended appearance in the FOCA frames nor in the optical DSS plates. However,
for the faintest of them some doubts arise from their optical appearance:
they could be very compact galaxies whose external faint
surface brightness envelopes are lost in the DSS plates.
Given that the spatial resolution of the FOCA experiment is
poor (FWHM $\approx 20$~arcsecs), a star/galaxy separation based on
standard methods is not possible. Thus a systematic search in the
USNO-B1.0 catalog (Monet et al. 2003) for the optical counterparts of
our UV sources was performed.
In order to shed light on the nature of these non extended UV sources, a diagnostic
based on the fluxes at 0.2$\mu$m (from FOCA), and at 0.44$\mu$m and
0.65$\mu$m (corresponding to the Johnson $B$ and Cousins $R$
respectively, from the USNO-B1.0 catalog) was applied. 
We present in Fig~\ref{br} the $AB_{0.2}-AB_{0.44}$
vs. $AB_{0.44}-AB_{0.65}$ diagram for our 375 UV sources and we
compare their combined UV-optical colors with those
of real galaxies and with the prediction of synthetic models of galaxies. 
We have selected an area in the figure which encloses the region
expected for galaxies,
taking into account the typical uncertainties of the USNO/FOCA 
photometry\footnote{Details
about the uncertainties of the USNO-B1.0 photometry are given in
Appendix~A}. The shaded area was defined from the position of the observed galaxies (observed with FAUST experiment) and was made larger using the typical photometric errors of galaxies in order to be
conservative.
Most of the UV sources, 314 in total, fall out of this  region
and thus they are discarded from the list of galaxy
candidates\footnote{Optical frames from the blue DSS plates centered at
these objects are available upon request: jorge.iglesias@oamp.fr}. \\
\item The remaining 61 UV sources, falling within the shaded region of Fig~2, have
colors consistent with those of real and
synthetic galaxies.
Blue DSS plates centered at the optical counterparts of these 61 UV
sources are presented in Fig~\ref{colorgal}. 
After a visual inspection of the optical plates we conclude that excepting 6, all of the sources are undoubtedly point-like objects and they are ruled out as being galaxy candidates.
\item The UV-optical colors of these 6 UV sources are listed in 
Table~\ref{nopoint}.
m030-768 and m028-817 are removed since they are listed as 
stars with spectral types
WD and sdB respectively\footnote{m028-817 is misclassified
as a cluster galaxy in SIMBAD based on the report of
Brosch et al. (1998), but in this paper the authors clearly classify this
object as the star TON~140.} in the SIMBAD database.
Thus, at the end only 4 UV sources remain as uncertain objects.\\
\item Out of the remaining four UV objects, three of them, m010-538, m018-542 and m033-310, are listed as point sources in the 2MASS All Sky Point Source Catalog. Although this fact enforces their
point-like nature, it is completely conclusive, being necessary multiwavelength or spectral
information to accurately determine their real nature.
None of these four objects have a FIR counterpart in the IRAS FSC catalog.
Since no evidence was found for them being galaxies,
we will not include these objects in the UVsel sample, we only
keep them as objects with an uncertain nature.\\
\end{enumerate}

At the end our UVsel sample is
composed by the 25 galaxies found after a crosscorrelation our list of UV
detections with the NED catalog. The selection of NED objects does not introduce a bias from the selection of the
dubious objects because the NED objects would have been classified as {\em secure} galaxies by just a visual inspection (as is evident from the optical plates presented in
Fig.~1). So, the only bias may come from the selection by-eye of objects from the DSS blue plates, in the sense that very compact objects are not included in the list of {\em secure} galaxies (but they are
listed as dubious objects).
In Table~\ref{UVseltab} we present some of the basic
properties of the 25 UV selected galaxies, as follows:

{\em Column (1)}: Galaxy name, from LEDA database.

{\em Column (2)}: FOCA field were the galaxy was detected.

{\em Column (3)}: Uncorrected UV ($0.2\mu$) magnitude in the AB system at 0.2$\mu$m.


{\em Column (4)}: Galactic extinction correction at 0.2$\mu$m according to Schlegel
et al. (1998) and the MW extinction curve of Pei (1992). 

{\em Column (5)}: Flux density at 60$\mu$m from the IRAS FSC catalog,
in Jy.

{\em Column (6)}: Flux density at 100$\mu$m from the IRAS FSC catalog,
in Jy.

{\em Column (7)}: $\log$ of the optical axial ratio, from the LEDA
database.

{\em Column (8)}: Morphological type, from NED database.

{\em Column (9)}: Distance to the galaxy in Mpc, from LEDA
database\footnote{The LEDA database assumes $H_{0} =
75$~km~s$^{-1}$~Mpc$^{-1}$}

{\em Column (10)}: $B$-band magnitude, from LEDA database.

{\em Column (11)}: Aggregation, from GOLDMINE (Gavazzi et al. 2003).

The sample galaxies are
mainly spirals (21 out of the total of 25) with only 3 irregulars and
one elliptical. 

The magnitudes used in the following analysis are
corrected for Galactic extinction using the values listed in
Table~\ref{UVseltab}. 

\subsection{Properties of the UVsel sample}

In this section we show some properties our UVsel sample and compare them with those of other UV samples of galaxies.

Fig~\ref{distuv} shows the distribution of distances of the UVsel galaxies. Two peaks are present in the distribution, due to the presence of cluster galaxies (mainly Virgo, Coma and Abell~1367). 
The median distance of the total sample is 56~Mpc. The shape of the distribution becomes flatter when we remove the cluster galaxies, and the median distance changes to 59~Mpc. 

In Fig~\ref{counts}, we present our UV galaxy counts corrected for
Galactic extinction compared to the UV galaxy counts brighter
than $AB_{0.2}=21$ 
already published: Milliard et al. (1992) with FOCA
data and Deharveng et al. (1994) with FAUST data. 
An extrapolation of the counts of Milliard et al. with slope 0.6,
representative of an euclidean distribution, has been
overplotted for comparison. 
In order to visualize the effect of the cluster galaxies we show both the total counts and those corresponding only to galaxies not associated to clusters.
The UV counts not including
the cluster galaxies agree well with the FAUST counts and deviate
significantly from the euclidean extrapolation at $AB_{0.2} \approx
16$~mag ($\approx 0.7$~dex). When including the cluster contribution, the total UV counts
increase, but even 
in this case the counts lie below the euclidean extrapolation to the
counts of Milliard et al. (1992).

Despite our poor statistics, we check whether
our UVsel sample is representative of the local 
Universe by comparing its UV LF, derived using the $1/V_{max}$ method\footnote{We call the attention of the reader to the fact that the $1/V_{max}$ method suffers from
systematic errors in the derived parameters of the LF for small samples of galaxies (see Willmer 1997 for details).}, 
to that of Sullivan et al. (2000), as shown in Fig~\ref{uvlf}.
It can be seen that our sample lacks of galaxies brighter
than $M^{AB}_{0.2} \simeq -17$ as compared to the sample of Sullivan et
al. whereas for fainter magnitudes both samples show a fairly good
agreement. When excluding the cluster galaxies of our sample, our LF shows a similar shape as that of Sullivan et al. and is shifted by $\approx 0.5$~dex at all UV magnitudes. This shift in the normalization of the LF is related to
the difference between our UV counts and the extrapolation 
to those of Milliard et al. 
We stress the fact that our goal in this work is not to construct an UV LF but just to compare the representativity of our UVsel sample to larger samples of UV
galaxies in the local Universe.

\subsection{FIR identifications \label{crossUV}}

We searched for FIR counterparts of our UVsel sample using the
60$\mu$m detections from the IRAS FSC catalog.
For each UV galaxy we selected a FIR
counterpart if the UV coordinates 
fall within the 3-$\sigma$ uncertainty
ellipse centered at the position of each FIR source. 
The uncertainty in the UV coordinates  ($\approx 2.5$~arcsecs)
was not taken into account since it is much smaller than the IRAS one
($\approx 40$~arcsecs for the major axis of the 1-$\sigma$ uncertainty ellipse). 

The FIR flux densities of the galaxies NGC~3883, NGC~4411
have been measured with the IRAS Scan Processing and
Integration facility (SCANPI) since no FIR sources associated to them
were present in the FSC catalog. 
At the end, 24 out of the 25 UV galaxies are identified at 60
$\mu$m, which corresponds to a detection rate of 96\%. 
The only galaxy without a FIR counterpart is the elliptical galaxy
NGC~4472. 

\section{The FIR selected sample}

\subsection{Selection of FIR galaxies}

The FIRsel sample is extracted from the
IRAS PSCz catalog.
The PSCz catalog was chosen  to simplify the cross identification with FOCA since this catalog, complete to $f_{60} = 0.6$~Jy, includes only galaxies with an optical identification.

42 galaxies from the PSCz fall within 
 our surveyed region of the sky. 
Some basic properties of the sample galaxies are
listed in Table~\ref{FIRseltab}.
The columns are the same as in Table~\ref{UVseltab} except 
for the identifier in column 2 which is now the PSCz identifier of the 
galaxies.

A comparison with the UVsel sample yields a total of 13 galaxies in common.

\subsection{Properties of the FIRsel sample}

As for the UV selected sample we first analyze the statistical properties of 
the FIRsel sample by comparing with those of other FIR selected samples.

We show the distribution of distances of the FIRsel galaxies in Fig~\ref{distfir}. It can be clearly seen that we are sampling a larger volume of the Universe than for the UVsel sample. Again the
distribution of distances shows two peaks at the redshifts of the clusters (approximately 17 and 90~Mpc). The average distance of the FIRsel sample is 81~Mpc, and it goes to 78~Mpc when removing the contribution of the cluster
galaxies, slightly higher than the average 45~Mpc of the total PSCz catalog (Saunders et al. 2000). We attribute this difference to the small statistics of our sample.

Fig~\ref{fircounts} shows the galaxy counts as a function of
$f_{60}$ for the FIRsel sample. 
Again we plot separately the total counts and those excluding the cluster galaxies. As a comparison we show the average counts for three different subsamples of the PSCz (Saunders et
al. 2000). Whereas for the total counts there is an excess
of galaxies at most flux densities, removing the contribution of the cluster galaxies reduces the differences with the PSCz counts.

We also show in Fig~\ref{firlf} the $L_{60}$ luminosity function of
the FIRsel sample estimated using the $1/V_{max}$ method, compared
with that of Takeuchi et al. (2003) for the PSCz using an analytical method. 
The LF of the total FIRsel sample is below the PSCz LF for galaxies with $10^{10} \leq L_{60} \leq 10^{11}~L_{\odot}$ and shows an excess for faint luminosities. The FIRsel LF restricted to galaxies
not associated to clusters is also below the LF for the total FIRsel sample and shows the two features mentioned above. 
The lack 
of galaxies with $10^{10} \leq L_{60} \leq 10^{11}~L_{\odot}$ is probably due to low statistics in both cases. However, the excess found at faint luminosities, although not present in the parametric
LF of Takeuchi et al. (2003), is 
also found by these authors when computing the LF using the $1/V_{max}$ method. 

\subsection{UV identifications}

We searched for UV counterparts to the FIRsel sources using the entire FOCA
dataset (down to $AB_{0.2} \simeq 20.3$) with a similar procedure as
the one detailed in Section~\ref{crossUV}. At the end of the
identification process, UV counterparts were found
for 38 of the FIRsel sources, which corresponds to a detection rate of
90\%.
For the four non detections -- Q02378+3829, Q09031+7855, Q13074+2852 and Q12217+0848 -- an upper limit to their UV flux corresponding to the detection limit of the corresponding frame is given.

\section{The Far-infrared to UV flux ratio and the internal extinction}

\subsection{The $F_{60}/F_{0.2}$ distribution}

We define $F_{60}$ and $F_{0.2}$ 
as $\lambda \cdot f_{\lambda}$ where  $f_{\lambda}$ is a
monochromatic density flux at 60$\mu$m and 0.2$\mu$m; the ratio
$F_{60}/F_{0.2}$ has no units. 
The histograms of $\log F_{60}/F_{0.2}$ are plotted in Fig~\ref{histfir_fuv} 
for both samples.
As expected the median values of $\log F_{60}/F_{0.2}$ are significantly different:
$0.19\pm 0.39$ for the UVsel sample
vs. $0.96\pm 0.59$ for the FIRsel
sample. When removing the cluster galaxies of both samples we obtain median values of 0.25 and 1.04 for the UVsel and FIRsel samples respectively, consistent with the values obtained for the total samples.
The differences in the median values obtained for $\log F_{60}/F_{0.2}$ 
can be understood in terms of selection biases,
as it is illustrated 
in Fig~\ref{plotfir_fuv}.
The FIR selection  (limited by the low horizontal line in the
Fig~\ref{plotfir_fuv}) favors galaxies with high
$F_{60}/F_{0.2}$ ratio. At
the opposite, the UV selection (limited by the right vertical line in the
Fig~\ref{plotfir_fuv}) is biased towards galaxies with low
$F_{60}/F_{0.2}$ values.

The very high detection rates found in UV
(respectively FIR) for the FIRsel (respectively UVsel) sample  as well
as the small difference in $F_{60}/F_{0.2}$ found between the
samples argue for a tight distribution of FIR/UV fluxes in galaxies. 
Nevertheless, because of our small statistics, we cannot
exclude the presence of different sub-populations of galaxies 
such as the very blue galaxies with very low values of
$F_{60}/F_{0.2}$ predicted by some models of galaxy evolution
(C. Xu, private communication). No such galaxies are present in our
UVsel sample\footnote{Although in the case that any of the four dubious objects listed in Section~3.1 were confirmed galaxies they would show a very low dust content}. 

The FIRsel sample contains four galaxies with no UV
counterpart. These galaxies, together with the few ones with $A_{0.2}>5$~mag,
might belong to a population of bright FIR
objects with a higher than normal $F_{60}/F_{0.2}$ ratio. This kind of
galaxies, already reported by several authors (e.g. Sanders \& Mirabel
1996; Trentham et al. 1999), are rare in the local
Universe (Buat et al. 1999).

\subsection{Internal dust extinction\label{extin}}

It is now commonly accepted that a reliable method to estimate the dust
extinction is to compare the emission of the dust to the emission of the stars 
observed  in UV.
The method is based on an energetic budget: 
all the stellar light absorbed 
by the dust is re-emitted in FIR-submm wavelengths. The models account for 
the dust heating by all the stars (old and young),  they assume a  
geometrical distribution for the dust and the stars  as well as   dust  
characteristics (absorption and scattering) and they solve the radiation 
transfer equation to deduce the extinction at all wavelengths.  The 
complexity of the analysis varies according to the authors  and their 
goals (e.g. Xu \& Buat 1995; Granato et al. 2000; Popescu et al. 2000; Panuzzo 
et al. 2003).
 Within the frame of these models, the comparison of the total dust emission 
and the observed UV one gives and estimate of the  extinction at UV 
wavelengths which is found very robust against the details of the models, as 
soon as the galaxies form stars actively (Buat \& Xu 1996; Meurer et al. 1999; 
Panuzzo et al. 2003; Gordon et al. 2001). 
 All the models need as input the total dust emission (10--1000$\mu$m).
This can be done by extrapolating the FIR flux to the total dust emission.
In general one starts with the FIR (40--120$\mu$m) flux 
computed as the combination of the fluxes at 60 and 100$\mu$m (Helou et al. 
1988) and applies a correcting factor $F_{\mbox{\scriptsize
dust}}/F_{\mbox{\scriptsize FIR}}$ (Xu \& Buat 1995; 
Meurer et al 1999; Calzetti et al. 2000).
Recent ISO data allowed an accurate
determination of $F_{\mbox{\scriptsize
dust}}/F_{\mbox{\scriptsize FIR}}$. Here we use the calibration proposed by 
Dale et
al. (2001) for $F_{\mbox{\scriptsize dust}}/F_{\mbox{\scriptsize FIR}}$ 
because it is suited for normal galaxies and only based on
the fluxes at 60 and 100$\mu$m. With this calibration we find
median values of $F_{\mbox{\scriptsize
dust}}/F_{\mbox{\scriptsize FIR}}$ of $2.4\pm1.3$ for the UVsel sample
and 2.3$\pm$1.1 for the FIRsel sample. The values obtained after removing the cluster galaxies are 2.41 and 2.29 for the UVsel and FIRsel samples respectively.
Following Buat et al. (1999) and adopting the new $F_{\mbox{\scriptsize dust}}/F_{\mbox{\scriptsize FIR}}$ calibration, the extinction can be expressed as
\begin{equation}
A_{0.2} = 0.622 + 1.140 x + 0.425 x^{2}
\end{equation}
where $x = \log F_{\mbox{\scriptsize FIR}}/F_{\mbox{\scriptsize
UV}}$. 
This formula is fully consistent with that used in HBI
(their Eq.~14 but expressed in terms $\log F_{\mbox{\scriptsize dust}}/F_{\mbox{\scriptsize UV}}$) since both are issued
from the same model. Here we prefer to use the observable quantities
$\log F_{\mbox{\scriptsize FIR}}/F_{\mbox{\scriptsize UV}}$.
An important issue for the following analysis is that there is no need to 
correct the dust  emission from the heating by old stars to derive the 
extinction from the FIR to UV flux ratio since the contribution of all the 
stars to the dust heating is accounted for by the models and included in the 
calibration of Eq.~1 which deals with the total FIR emission.
Fig~\ref{compa_extin} compares the
calibration of $A_{0.2}$ given in Eq.~1 with the ones of Buat et al. (1999), Calzetti et al. (2000) and Panuzzo et al. (2003), illustrating that the
calibration proposed in this work (Eq.~1) is, in particular, in
very good agreement with the one proposed by Panuzzo et al. (2003) for face-on
galaxies. These authors take into account the inclination
angle of the galaxy when computing the extinction. They also propose  a
calibration for the extinction of nearly edge-on galaxies which results
in higher extinctions for the same value of $F_{\mbox{\scriptsize
FIR}}/F_{\mbox{\scriptsize UV}}$. 

When applying Eq~1 to our samples, we obtain median values of $A_{0.2}$ of
$0.98\pm 0.54$~mag and $2.20\pm1.21$~mag for the UVsel and the
FIRsel
samples respectively. Removing the cluster galaxies yields median values of 1.00 and 2.38 for the UVsel and FIRsel samples respectively, again consistent with the values obtained for the total samples.
The median value of $A_{0.2}\approx 1~$mag found for the UVsel sample is
consistent with the one reported by Buat \& Xu (1996) -- $A_{0.2} =
0.9$~mag -- for a 
sample of galaxies essentially UV selected: it was based on UV
detections from the SCAP (Donas et al. 1987) and FAUST (Deharveng et
al. 1994) observations with available FIR counterparts.
The value obtained for the FIRsel sample is larger than the one
reported by Buat et al. (1999) -- $A_{0.2} = 1.6$~mag -- for a sample
of galaxies with UV and FIR data in the FOCA and IRAS FSC catalogs
respectively. However, this last sample was not truly FIR selected 
since it only contained FIR detections confirmed as galaxies in the IRAS FSC catalog of associations and thus some galaxies with a FIR counterpart but with an association of a different nature in this
catalog were rejected.

Concerning the four galaxies with no UV counterpart, their $L_{60}$
as well as a lower limit for $A_{0.2}$ are given  in Table~\ref{threeFIR}. 
These four galaxies show very large IR luminosities
(two of them are LIRGs with $L_{60}> 10^{11}
L_{\odot}$). They are also detected as radio sources by the NVSS survey.
This kind of galaxy, often associated to interacting systems (Sanders \& Mirabel
1996), is not very common in the local Universe. In fact, this could be the
case of Q02378+3829, whose optical counterpart reveals the existence
of a close pair of galaxies. 
Values of the extinction of $A_{0.2} \simeq 6.5$~mag have been
reported for nearby LIRGs from other samples in the literature (Buat
et al. 1999), consistent with $A_{0.2} \geq 3.8$~mag we find
for these four galaxies.

\section{Calibration of the FIR and UV emissions as SFR tracers}

\subsection{SFR estimators}

Both UV and FIR emissions in star forming galaxies are related to young 
stars and thus are potential tracers of the recent star formation. 
The aim of this section is to make use of the
FIR and UV emissions to properly estimate the SFRs for our
UVsel and FIRsel samples and to propose a method to obtain reliable
SFRs when only UV or FIR fluxes are available.

Various assumptions must be made for the calculation of the SFR from
UV or FIR luminosities. A star formation history and an initial mass
function (IMF) must be assumed in order to relate the stellar emission
to the SFR. Hereafter we make the assumption of a constant
SFR in the last $\simeq 10^{8}$~yr and we
adopt a Salpeter IMF between 0.1 and 100$M_{\odot}$ in order to
convert UV and FIR luminosities to SFRs.
Under these conditions, the conversion between the luminosity of a
galaxy at a given wavelength and its SFR is given by stellar synthesis 
models. Here we list the most popular methods to estimate the SFRs of
galaxies from UV and/or FIR luminosities.

\begin{enumerate}
\item {\bf SFR deduced from the UV flux corrected for dust extinction.}
The assumption of a constant SFR in the last $10^{8}$~yr together with
the fact that most of the UV flux comes from stars with lifetimes
$\leq 3 \times 10^{8}$~yr allows to estimate the SFR from the UV
luminosity. The validity of this approximation requires a
proper correction for internal extinction at UV wavelengths. 
This method has been used intensively (Donas et al. 
1987; Buat \& Xu 1996; Buat et al. 2002; Madau et al. 1998; Bell \&
Kennicutt 2001; Boselli et al. 2001). 
Using the code Starburst99
with a Salpeter IMF from 0.1 to 100 solar masses we obtain:
\begin{equation}
SFR_{\mbox{\scriptsize UV}} = 2.03\times 10^{-40}~L_{\mbox{\scriptsize 
0.2}}^{corr}
\label{uvsfr}
\end{equation}
with  $SFR_{\mbox{\scriptsize UV}}$ in $M_{\odot}$~yr$^{-1}$ and 
$L_{\mbox{\scriptsize
0.2}}^{corr}$ in erg~s$^{-1}$~\AA$^{-1}$. $L_{\mbox{\scriptsize
0.2}}^{corr}$ corresponds to the UV luminosity at 0.2$\mu$m corrected for 
dust extinction using Eq~1. This calibration is consistent with that of Kennicutt (1998) within less than 10\%.\\

\item {\bf SFR is calculated from the dust emission.}
This method 
relies on two major assumptions: (1) the dust extinction must be
high enough in order that the stellar light escaping the galaxy is
negligible and (2) the contribution of the old
stars to the dust heating is negligible. These assumptions are likely to be
true for galaxies very active in star forming and FIR luminous. 
However, the extension of this approximation to normal star formation galaxies (e.g. optically selected) becomes problematic due
to the contamination of the FIR emission by a cooler cirrus component heated by stars older than $\approx 10^{8}$~yr (Sauvage \& Thuan 1992), and also because of the stellar light which escapes the
galaxies without being absorbed by the dust.
A further difficulty of this method is to estimate the total dust emission from 
the observations: they are often restricted to 60 and/or 100$\mu$m and
average factors must be used to obtain the $F_{\mbox{\scriptsize
dust}}/F_{\mbox{\scriptsize FIR}}$ ratio. We apply the $F_{\mbox{\scriptsize
dust}}/F_{\mbox{\scriptsize FIR}}$ calibration from Dale et al. (2001, c.f. Sec.~5.2) and then we obtain the $SFR_{\mbox{\scriptsize dust}}$ following HBI:
\begin{equation}
SFR_{\mbox{\scriptsize dust}} = 4.7\times
10^{-44}~L_{\mbox{\scriptsize dust}}
\label{firsfr}
\end{equation}
with $SFR_{\mbox{\scriptsize dust}}$ in $M_{\odot}$~yr$^{-1}$ and
$L_{\mbox{\scriptsize dust}}$ in erg~s$^{-1}$. Again this calibration is consistent with that of Kennicutt (1998) within less than 5\%.\\

\item {\bf SFR is deduced by combining the UV 
(uncorrected for extinction) and the dust emissions.}
This method takes into account the UV light coming directly from young stars as well as the emission of the dust heated by these young stars. 
The observed UV flux and the dust emission at FIR
wavelengths can be calibrated into SFR using the above expressions
(Eqs.~\ref{uvsfr} and \ref{firsfr}):
\begin{displaymath}
SFR_{\mbox{\scriptsize dust$+$UV}} =SFR_{\mbox{\scriptsize
dust}} + SFR_{\mbox{\scriptsize UV}}^{obs} =
\end{displaymath}
\begin{equation}
= 4.7\times 10^{-44}~L_{\mbox{\scriptsize dust}}+2.03\times
10^{-40}~L_{\mbox{\scriptsize 0.2}}^{obs}
\label{sfrtot}
\end{equation}
where $L_{\mbox{\scriptsize dust}}$ is expressed in erg~s$^{-1}$,
$L_{\mbox{\scriptsize 0.2}}^{obs}$ in erg~s$^{-1}$~\AA$^{-1}$ and the SFR
in $M_{\odot}$~yr$^{-1}$.\\

As previously quoted, in normal star forming galaxies (not starbursting) a fraction of FIR luminosity is due to old stars.
To properly estimate the SFR, this
contribution should be removed from the conversion formula. 
Under this assumption, Eq~\ref{sfrtot} is modified as follows:
\begin{displaymath}
SFR(\eta) = SFR_{\mbox{\scriptsize
dust}}(\eta) + SFR_{\mbox{\scriptsize UV}}^{obs} =
\end{displaymath}
\begin{equation}
= (1-\eta) \times 4.7\times 10^{-44}~L_{\mbox{\scriptsize dust}}+2.03\times
10^{-40}~L_{\mbox{\scriptsize 0.2}}^{obs}
\end{equation}

HBI have recently studied the consistency of the 
H$\alpha$, UV and FIR emissions of
galaxies in terms of SFR in order to estimate the fraction
of dust heated by stars older than $10^8$~yr. 
They found $\left<\eta\right> = 0.4 (\sigma = 0.06)$ for a sample of star
forming galaxies selected to be observed in UV, FIR and in the Balmer (at least H$\alpha$) emission lines. The $\eta$ parameter was found to be robust against variations of model
assumptions: SFR constant over $10^{7}$ or $10^{8}$~yr, or slight variations of $L_{\mbox{\scriptsize dust}}/L_{\mbox{\scriptsize FIR}}$. Such a low dispersion found for $\eta$ is probably due to the
selection of the sample, which contains optically selected nearby spiral or irregular galaxies active in star formation. Indeed for a sample of starburst galaxies UV selected, HBI found $\eta \approx
0$, all the dust heating being attributed to young ($\leq 10^{8}$~yr) stars.
A very low value of $\eta$ is also probably valid for galaxies whose spectral energy distribution is dominated by their FIR luminosity, with very intense star formation activity.
Nevertheless, as shown in Fig.~11, galaxies with a very high dust content are
almost absent of our samples: only four of the FIRsel galaxies show extinction larger than $A_{0.2} = 4.6$~mag. For two of them only an upper limit to the UV flux is available
(and thus, they could present a very large extinction) and they will not be included in the
subsequent analysis.
Xu et al. (1994) also estimated the contribution to the FIR ($40 - 120\mu$) emission of stars less massive than
$5M_{\odot}$ in spiral galaxies and found that $\approx 30$\% of the FIR emission is due to these stars. More recently, Misiriotis et al. (2001) gave an upper limit of $\approx 40$\% for the heating
by old stars distributed in a thick disk and the bulge. One of the aims of the present paper is to test whether the value of $\eta$ estimated in HBI for star forming galaxies remains valid, at least
on average, for galaxy samples selected according to different criteria.
\end{enumerate}

Using $SFR(\eta)$  or $SFR_{\mbox{\scriptsize dust$+$UV}}$ as a quantitative 
estimator of the SFR assumes implicitly
that both the observed UV and FIR emission are isotropic since they relate the 
luminosities to SFR. 
However the UV emission of a galaxy affected by the extinction is 
certainly 
not isotropic. Therefore the relations defined by Eqs~4 and 5 have
only a statistical significance assuming that the galaxies are
randomly oriented. This caveat is less important for
$SFR_{\mbox{\scriptsize
UV}}$ and $SFR_{\mbox{\scriptsize dust}}$ since the dust and UV emissions
corrected for dust extinction are supposed to be isotropic.

In the following we will compare these four determinations of the SFR
-- $SFR_{\mbox{\scriptsize UV}}$, $SFR_{\mbox{\scriptsize dust}}$,
$SFR_{\mbox{\scriptsize dust$+$UV}}$ and $SFR(\eta)$ -- for both our
UVsel and FIRsel samples with the aim of comparing the results of the
different methods and of finding the best way to
estimate the SFR when only UV or FIR data are available. 

\subsection{Estimating the SFR in the UVsel sample}

Fig~\ref{sfr_UVsel} shows the comparison between the four
different estimations of the SFR for the galaxies in the UVsel sample (excepting the galaxy for which only un upper limit to the FIR flux is available). 

In panel~{\bf a} we present the comparison between the SFRs estimated from the
UV fluxes using a proper extinction correction for each galaxy
($SFR_{\mbox{\scriptsize UV}}$) and those estimated also from the UV
fluxes but using the average extinction correction
$A_{0.2}=1$~mag, a value similar to the one found for our UVsel
sample. This last estimation of the SFR is
very useful for UV selected samples when FIR data are not
available. It appears that the agreement
between both determinations of the SFR is quite good over two orders
of magnitude in SFR.

The remaining three panels of the figure show the comparison between
$SFR_{\mbox{\scriptsize UV}}$ and the other three estimators 
mentioned above.
As it is shown in panel~{\bf b}, $SFR_{\mbox{\scriptsize UV}}$ is
systematically lower than $SFR_{\mbox{\scriptsize
dust+UV}}$ ($\approx 0.14$~dex). Panel~{\bf c} shows that using
$SFR(\eta)$ with $\eta=0.4$ leads to an
almost perfect agreement between $SFR_{\mbox{\scriptsize UV}}$ and
$SFR(\eta)$, thus confirming the conclusion reached by HBI.
Finally, we show in panel~{\bf d} that $SFR_{\mbox{\scriptsize dust}}$ is in quite
good agreement with $SFR_{\mbox{\scriptsize UV}}$. 
This result comes out from the combination of two effects (dust heating due to old stars is not subtracted and the contribution of the young stars not absorbed by the dust is not included in the
$SFR_{\mbox{\scriptsize dust}}$
estimation) which are not taken into account but which compensate each other producing the agreement between $SFR_{\mbox{\scriptsize dust}}$ and $SFR_{\mbox{\scriptsize UV}}$. We stress however the
accidental nature of this relationship.

In general, we show that the slopes of all these relations are close
to unity, that is, the shifts between the different estimations of the SFR
are almost constant over the whole range of SFRs of our UVsel sample
and no trends with the luminosity are found.

These results seem to be in conflict with the trend found by HBI 
(see also Buat 2003) that the dust emission under-predicts the SFR for 
galaxies with SFR~$< 1 M_{\odot}$/yr. Once again the difference is probably due to 
selection effects: the galaxy sample used by HBI is biased 
towards late type galaxies for which a Balmer decrement (and therefore a 
H$\beta$ line) has been measured and does not  sample accurately the UV 
luminosity function. Indeed the mean extinction found by HBI 
is only 0.75 mag against 1 mag for the present UVsel sample.

\subsection{Estimating the SFR in the FIRsel sample}

Concerning the FIRsel sample, we show in Fig~\ref{sfr_firsel} the
comparisons between $SFR_{\mbox{\scriptsize dust}}$ and the other 
 estimators already proposed: $SFR_{\mbox{\scriptsize UV}}$,
$SFR_{\mbox{\scriptsize dust+UV}}$, $SFR(\eta)$. Only FIRsel galaxies detected in the UV frames are included in this analysis.

$SFR_{\mbox{\scriptsize dust}}$ is almost
similar to $SFR_{\mbox{\scriptsize dust+UV}}$, as shown in panel~{\bf a}, since the
contribution of the observed UV flux is rather negligible for the FIRsel galaxies. 
In the same way, decontaminating from the dust heated by old stars 
leads to $SFR(\eta) \sim 60$\% lower
than $SFR_{\mbox{\scriptsize dust+UV}}$ and therefore than
$SFR_{\mbox{\scriptsize dust}}$, as it is illustrated in panel~{\bf b}. 
Contrary to what observed in the UV selected sample,
$SFR_{\mbox{\scriptsize UV}}$ is lower than 
$SFR_{\mbox{\scriptsize dust}}$, as illustrated in panel~{\bf c}. 
This different behavior is probably 
due to the fact that the accidental compensation between the cold
stellar contribution to the total dust emission and the
emission of the young stellar population not absorbed by dust, observed in the UVsel
sample, is here not reproduced.
Finally, panel~{\bf d} shows that
$SFR_{\mbox{\scriptsize UV}}$ is in good agreement with $SFR(\eta)$ as
it was also found for the UVsel sample. Combining the results from
panels~{\bf c} and {\bf d} we argue that the assumption that all the dust
emission comes from young stars is not valid for galaxies 
like the ones we find in our samples, 
and that we must assume that a fraction of this emission
comes from older stars. 

The slopes of the relations between the SFRs for the FIRsel samples are 
again
close to unity meaning that no trend with luminosity does exist and
that the conversions between the different methods to estimate the SFR
are valid over the range of SFRs covered by our sample. 
Even the two galaxies included in this analysis with $A_{0.2} \geq 4.6$~mag follow the same trends as the other galaxies of the FIRsel sample. This fact
argues for the validity of the calibrations between the different estimators of the SFR even for very extincted galaxies, expected to be more numerous at high $z$. However, larger and deeper samples
are necessary to confirm this hypothesis.

\subsection{SFR vs. extinction for UVsel and FIRsel samples}

It has been reported by several authors a correlation between the
extinction and the SFR (Calzetti et al. 1995; Wang \&
Heckman 1996; Buat et al. 1999; Adelberger \& Steidel 2000; 
Hopkins et al 2001; Sullivan et al
2001), which may be attributed to both the increase of metallicity 
and the increase of surface mass density with increasing
luminosity (Wang \& Heckman 1996). In order to verify this effect, 
we show in Fig~\ref{sfrtot_fdustfuv} $A_{0.2}$ versus 
$SFR(\eta)$ for the UVsel and FIRsel samples. A very
dispersed correlation is apparent between both quantities when
considering both samples together. However this trend seems to be
mainly due to the FIRsel galaxies since it disappears when
considering only the UVsel galaxies. In any case, it is clear
that the scatter of the relationship is enhanced by the different
selection criteria. 

We show in Fig~\ref{sfrtot_s_fdustfuv} $A_{0.2}$ versus 
$SFR(\eta)/Area$, where $Area$ is the optical surface of the galaxies in kpc$^{2}$. As can be seen, a very dispersed correlation is apparent both samples although the larger
scatter corresponds again to the FIRsel sample.

\section{Conclusions}

We have presented the properties of two samples of galaxies selected
from the same region of the sky by their UV and FIR fluxes. The
detection rate for the UVsel sample at FIR wavelengths was found to be 96\% whereas that for the FIRsel sample at UV wavelengths equals 90\%.
The UV counts are lower than the expected from the extrapolation of previous determinations at fainter magnitudes, even when including the contribution of the cluster galaxies.
We showed that their dust extinction properties are different, the UV
selected galaxies exhibiting a lower extinction than the FIR selected
ones ($\approx$1~mag on average for the UVsel vs. $\approx$2~mag
for the FIRsel). 
Four galaxies of the FIRsel sample do not have a UV counterpart,
implying lower limits to the UV extinction of $3.8$~mag. These
galaxies could be part of the population of very extincted objects
already reported in the literature.

We compared  different indicators of the SFR calculated with the FIR and/or 
UV  luminosities and we
showed that they correlate well with each other for both samples.
The relations between the different
estimators of the SFR present a slope close to unity for both samples,
meaning that 
no trend with the SFR exists when converting between each other.
For both samples we found the best agreement between the following quantities: (a) the SFR calculated from the UV luminosities corrected for dust extinction using the FIR/UV ratio and (b) the
sum of the SFR calculated from the dust luminosities corrected for the average contribution of the dust heating due to old stars ($\approx 40$\%) and of the SFR calculated from the observed UV luminosities.

Putting both samples together we find the correlation  between
SFR and extinction already reported for other samples of galaxies but
with a very large scatter. Most of the trend is due to the galaxies selected in 
FIR.

The results of this work seem not affected by the cluster environment since we have shown that the global properties of cluster and field galaxies present in our samples are similar. This is
expected from the similarities between field and cluster LFs at UV and FIR wavelengths (Bicay \& Giovanelli 1987; Cortese et al. 2003).

\begin{acknowledgements}
Thanks are given to C. Xu for interesting suggestions and comments.
This research has made use of the NASA/IPAC Extragalactic Database
(NED) and the NASA/IPAC Infrared Science Archive, which are operated
by the Jet Propulsion Laboratory, California Institute of Technology,
under contract with the National Aeronautics and Space Administration.
This research has made use of the
SIMBAD database, operated at CDS, Strasbourg, France. 
The LEDA database (http://leda.univ-lyon1.fr/) was used throughout this work.
\end{acknowledgements}

\onecolumn

\clearpage

   \begin{table}
      \caption[]{Basic properties of the selected FOCA fields.}
         \label{tabla}
     $$ 
         \begin{tabular}{llccr}
            \hline
            \noalign{\smallskip}
FOCA field & Id. FOCA & R.A. (J2000) & Dec. (J2000) &
$AB_{0.2}^{lim}$ \\
            \noalign{\smallskip}
            \hline
            \noalign{\smallskip}
m015 & NGC 1023   & 02:36:41.8 & $+$38:43:11 & 19.36 \\
m010 & Cancer     & 08:20:24.6 & $+$20:45:07 & 20.36 \\
m018 & NGC 2715   & 08:58:30.4 & $+$78:09:29 & 20.06 \\
m067 & Abell 1367 & 11:45:24.5 & $+$19:11:39 & 20.26 \\
m033 & NGC 4125   & 12:01:27.1 & $+$64:56:17 & 20.56 \\
m050 & NGC 4472   & 12:27:38.4 & $+$08:35:57 & 19.86 \\
m028 & Coma       & 12:59:37.0 & $+$28:06:29 & 21.06 \\
m030 & SA 57a     & 13:06:15.4 & $+$29:04:20 & 21.06 \\
m031 & SA 57b     & 13:11:50.1 & $+$27:52:24 & 20.46 \\
            \noalign{\smallskip}
            \hline
         \end{tabular}
     $$ 
  \end{table}

\newpage
\clearpage

\begin{table}
\caption[]{Photometric properties of the UV sources with uncertain
nature. Data are not corrected for Galactic extinction.}
\label{nopoint}
$$
\begin{tabular}{lccc}
\hline
\noalign{\smallskip}
FOCA id. & $AB_{0.2}$ & $AB_{0.2} - AB_{0.44}$ & $AB_{0.44} -
AB_{0.65}$ \\
 & (mag) & (mag) & (mag) \\
\noalign{\smallskip}
\hline
\noalign{\smallskip}
m010-538 & 16.71 & 3.80 & 0.66 \\
m010-113 & 16.80 & 0.62 & -0.76 \\
m018-542 & 15.90 & 0.51 & -0.85 \\
m028-817 & 14.99 & 0.12 & -0.99 \\
m030-768 & 13.93 & -0.87 & -0.64 \\
m033-310 & 16.06 & 1.23 & 0.37 \\
\noalign{\smallskip}
\hline
\end{tabular}
$$
\end{table}

\clearpage

\begin{table}
\caption[]{Basic properties of the UV selected sample galaxies}
\label{UVseltab}
$$
\begin{tabular}{lcrrrrrrrrr}
\hline
\noalign{\smallskip}
Name & FOCA field & $AB_{0.2}$ & $A^{Gal}_{0.2}$ & $f_{60}$ & $f_{100}$ & $\log 
D_{25}$ &
Type & Dist & $B_{T}$ & Agg.\\
     &            & (mag)      & (mag) & (Jy)     & (Jy)      &
&    & (Mpc) & (mag) & \\
\noalign{\smallskip}
\hline
\noalign{\smallskip}
   NGC~4848     & m028 &  16.29 & 0.08 &  1.34 &   2.60 &   1.18 &   SBab: & 
104.23 & 14.42 & Coma \\
   UGC~6697     & m067 &  15.41 & 0.19 &  1.52 &   2.88 &   1.27 &   Im: &  
97.72 & 14.24 & A1367 \\
   NGC~3861$^{\dagger}$     & m067 &  16.27 & 0.29 &  0.44 &   1.66 &   1.32 &  
 (R')SAB(r)b &  74.13 & 13.72 & A1367 \\
   NGC~3883     & m067 &  16.27 & 0.25 & 0.37 &   1.30 &   1.42 &   SA(rs)b & 
101.86 & 13.86 & Field \\
   CGCG~097-068 & m067 &  16.95 & 0.19 & 1.87 &   3.91 &   1.00 &   Sbc &  
86.70 & 14.75 & A1367 \\
   CGCG~097-079 & m067 &  16.98 & 0.21 &  0.33 &   0.64 &   0.81 &  Irr & 
101.86 & 16.14 & A1367 \\
   UGC~6743     & m067 &  16.97 & 0.21 &  0.39 &   0.70 &   1.13 &   SABbc &  
98.17 & 14.37 & Field \\
   UGC~4329     & m010 &  16.09 & 0.48 &  0.49 &   1.37 &   1.23 &   SA(r)cd &  
59.16 & 14.38 & Cancer \\
   CGCG~119-047 & m010 &  16.99 & 0.38 &  0.99 &   2.27 &   0.91 &   Sab &  
65.16 & 15.13 & Cancer \\
   IC~239       & m015 &  15.15 & 0.63 &  0.72 &   5.07 &   1.70 &   SAB(rs)cd 
&  14.72 & 11.81 & Field \\
   UGC~2069     & m015 &  15.97 & 0.44 &  1.19 &   2.93 &   1.31 &   SAB(s)d &  
55.72 & 14.48 & Field \\
   NGC~2715     & m018 &  14.68 & 0.23 &  1.84 &   1.02 &   1.68 &   SAB(rs)c & 
 22.70 & 11.90 & Field \\
   NGC~2591     & m018 &  16.36 & 0.19 &  1.63 &   5.28 &   1.48 &   Scd: &  
22.70 & 13.47 & Field \\
   VV~841       & m030 &  16.60 & 0.11 &  0.30 &   0.53 &   0.93 &   Irr &  
70.47 & 15.68 & Coma \\
   NGC~5000$^{\dagger}$     & m030 &  16.58 & 0.08 &  0.96 &   2.39 &   1.16 &  
 SB(rs)bc &  82.79 & 14.04 & Field \\
   CGCG~160-128 & m030 &  16.75 & 0.11 &  0.23 &   0.49 &   0.77 &   Sb & 
117.49 & 15.88 & Coma \\
   NGC~4470     & m050 &  15.25 & 0.21 &  1.86 &   1.82 &   1.12 &   Sa? &  
34.36 & 13.02 & Virgo \\
   NGC~4411b    & m050 &  15.24 & 0.27 &  0.40 &   1.78 &   1.39 &   SAB(s)cd & 
 19.14 & 13.24 & Virgo \\
   NGC~4416     & m050 &  15.76 & 0.23 &  0.93 &   2.70 &   1.21 &   SB(rs)cd: 
&  20.80 & 13.24 & Virgo \\
   UGC~7590     & m050 &  16.01 & 0.19 &  0.38 &   0.92 &   1.10 &   Sbc &  
16.98 & 14.44 & Virgo \\
   NGC~4411     & m050 &  15.97 & 0.23 &  0.20 &   0.70 &   1.28 &   SB(rs)c &  
19.32 & 13.73 & Virgo \\
   NGC~4472     & m050 &  16.04 & 0.21 & $<$0.20 & $<$0.80 &   1.99 &   E2 &  
13.43 &  9.28 & Virgo \\
   NGC~4424     & m050 &  16.29 & 0.19 &  3.31 &   5.92 &   1.53 &   SB(s)a &   
7.35 & 12.46 & Virgo \\
   NGC~4451     & m050 &  16.42 & 0.17 &  1.68 &   5.17 &   1.13 &   Sbc: &  
13.43 & 13.30 & Virgo \\
   NGC~4492     & m050 &  16.91 & 0.23 &  0.25 &   1.20 &   1.27 &  
 SA(s)a? &  26.18 & 13.22 & Virgo \\
\noalign{\smallskip}
\hline
\end{tabular}
$$
$\dagger$ Pair of galaxies not resolved by FOCA.
\end{table}

\clearpage

\begin{table}
\caption[]{Basic properties of the FIR selected sample galaxies}
\label{FIRseltab}
$$
\begin{tabular}{llrrrrrrrrrr}
\hline
\noalign{\smallskip}
Name & PSCz id. & $AB_{0.2}$ & $A^{Gal}_{0.2}$ & $f_{60}$ & $f_{100}$ & $\log 
D_{25}$ 
&
Type & Dist & $B_{T}$ & Agg. \\
     &          & (mag)      & (mag) & (Jy)     & (Jy)      &               & 
     & (Mpc)& (mag) & \\
\noalign{\smallskip}
\hline
\noalign{\smallskip}
   NGC~3860     & Q11422+2003 &   18.79 & 0.21 &  0.71 &   2.49 &   1.07 &   Sa 
& 81.28 &  14.32 & A1367 \\
   UGC~6697     & Q11412+2014 &   15.41 & 0.19 &  1.52 &   3.16 &   1.27 &   
Im: & 97.72 &  14.24 & A1367 \\
   NGC~3840     & Q11413+2021 &   17.33 & 0.21 &  0.82 &   1.78 &   0.99 &   Sa 
&106.66 &  14.79 & A1367 \\
   NGC~3859     & Q11423+1943 &   17.98 & 0.21 &  1.00 &   2.27 &   1.05 &   
Irr & 79.80 &  14.89 & A1367 \\
   IC~732$^{\dagger}$       & R11433+2043 &   19.87 & 0.21 &  3.43 &   6.08 &   
0.81 &   Pair &105.68 &  15.82 & A1367 \\
   CGCG~097-068 & Q11398+2023 &   16.95 & 0.19 &  1.82 &   4.02 &   1.00 &   
Sbc & 86.70 &  15.35 & A1367 \\
   CGCG~127-049 & R11432+2054 &   18.11 & 0.19 &  0.64 &   1.31 &   0.92 &  S 
&102.33 &  15.35 & A1367 \\
   IC~4040      & Q12582+2819 &   17.33 & 0.11 &  1.23 &   2.69 &   0.92 &   
Sdm: &114.82 &  15.33 & Coma \\
   KUG~1256+285 & Q12561+2832 &   18.67 & 0.11 &  0.75 &   0.81 &   0.32 &  S 
&420.73 &  17.38 & Field \\
   NGC~4848     & Q12556+2830 &   16.29 & 0.08 &  1.34 &   2.90 &   1.18 &   
SBab: &104.23 &  14.42 & Coma \\
   NGC~4911$^{\dagger}$     & Q12584+2803 &   17.16 & 0.08 &  0.72 &   2.50 &   
1.13 &   SAB(r)bc &116.41 &  13.71 & Coma \\
   NGC~4853     & Q12561+2752 &   18.04 & 0.08 &  0.64 &   1.55 &   0.91 &  
(R')SA0 &111.17 &  14.46 & Coma \\
   NGC~4926A    & Q12596+2755 &   18.05 & 0.08 &  0.64 &   1.11 &   0.78 &  S0 
&103.28 &  15.60 & Coma \\
   MRK~53       & Q12536+2756 &   17.34 & 0.08 &  0.63 &   1.68 &   0.50 &  Sa 
& 73.45 &  15.80 & Coma \\
   KUG~1300+276 & Q13008+2736 &   18.38 & 0.08 &  0.71 &   1.43 &   0.75 & S 
&152.05 &  16.26 & Field \\
   UGC~4332     & Q08166+2116 &   19.92 & 0.46 &  0.88 &   2.09 &   1.09 &   
Irr & 78.70 &  14.89 & Cancer \\
   UGC~4324     & Q08155+2055 &   19.29 & 0.38 &  0.61 &   1.60 &   1.05 &   
Sab: & 69.18 &  15.03 & Cancer \\
   UGC~4386     & Q08211+2111 &   18.86 & 0.36 &  0.66 &   2.65 &   1.27 &   Sb 
& 66.99 &  14.21 & Cancer \\
   IC~2339      & Q08206+2130 &   16.57 & 0.42 &  1.65 &   2.88 &   0.94 &   
SB(s)c & 77.98 &  14.98 & Cancer \\
   CGCG~119-047 & Q08161+2156 &   16.99 & 0.38 &  0.99 &   2.27 &   0.91 &   
Sab & 65.16 &  15.13 & Cancer \\
   IC~239       & Q02333+3845 &   15.15 & 0.63 &  0.72 &   5.07 &   1.70 &   
SAB(rs)cd & 14.72 &  11.81 & Field \\
   CGCG~523-086 & Q02378+3829 &$>$19.36 & 0.53 &  3.12 &   4.41 &    --- &  --- 
&216.77 & 15.69 & Field \\
   UGC~2069     & Q02325+3725 &   15.97 & 0.44 &  1.28 &   3.26 &   1.31 &   
SAB(s)d & 55.72 &  14.48 & Field \\
   NGC~2655     & Q08491+7824 &   17.52 & 0.29 &  1.53 &   5.09 &   1.68 &   
SAB(s)0/a & 23.77 &  11.16 & Field \\
   NGC~2715     & Q09018+7817 &   14.68 & 0.23 &  1.73 &   9.70 &   1.68 &   
SAB(rs)c & 22.70 &  11.90 & Field \\
   LEDA~139162  & Q09031+7855 &$>$20.06 & 0.17 &  0.64 &   1.11 &    --- &  --- 
&599.79 &   --- & Field \\
   NGC~2591     & Q08307+7811 &   16.36 & 0.19 &  1.69 &   5.17 &   1.48 &   
Scd & 22.70 &  13.47 & Field \\
   CGCG~160-151  & Q13068+2937 &   17.46 & 0.11 &  0.61 &   1.27 &   0.56 &   Sb 
& 92.04 &  15.27 & Field \\
   NGC~5000$^{\dagger}$     & Q13073+2910 &   16.58 & 0.08 &  0.92 &   2.45 &   
1.16 &   SB(rs)bc & 82.79 &  14.04 & Field \\
   Q13074+2852 & Q13074+2852 &$>$21.06 & 0.06 &  1.16 &   1.67 &    --- &---& 
92.04 &  --- & Coma \\
   NGC~4922$^{\dagger}$     & Q12590+2934 &   18.41 & 0.11 &  6.61 &   7.08 &   
1.10 &  Pair &103.75 &  13.89 & Coma \\
   CGCG~160-161 & Q13100+2848 &   17.55 & 0.08 &  1.36 &   2.02 &   0.66 &  S0 
&101.39 &  15.46 & Field \\
   VV~474       & Q13105+2724 &   18.52 & 0.11 &  0.78 &   1.71 &   0.94 &   
--- &100.46 &  15.67 & Field \\
   UGC~7020A    & Q12000+6439 &   17.24 & 0.19 &  1.79 &   2.63 &   1.08 &  S0? 
& 25.35 &  14.46 & Field \\
   NGC~4081     & Q12020+6442 &   17.98 & 0.19 &  1.89 &   4.24 &   1.18 &  Sa? 
& 24.43 &  13.73 & Field \\
   NGC~4125     & Q12055+6527 &   17.65 & 0.17 &  0.62 &   1.27 &   1.78 &  E6 
& 23.23 &  10.63 & Field \\
   NGC~4469     & Q12269+0901 &   17.29 & 0.19 &  1.16 &   3.05 &   1.54 &   
SB(s)0/a &  9.29 &  12.36 & Virgo \\
   NGC~4451     & Q12260+0932 &   16.42 & 0.17 &  1.71 &   4.49 &   1.13 &   
Sbc: & 13.43 &  13.30 & Virgo \\
   NGC~4416     & Q12242+0811 &   15.76 & 0.23 &  0.98 &   2.90 &   1.21 &   
SB(rs)cd: & 20.80 &  13.24 & Virgo \\
   NGC~4424     & Q12246+0941 &   16.29 & 0.17 &  3.21 &   6.07 &   1.53 &   
SB(s)a: &  7.35 &  12.46 & Virgo \\
   NGC~4356     & Q12217+0848 &$>$19.86 & 0.25 &  0.61 &   1.56 &   1.41 &   Sc 
& 17.22 &  14.04 & Virgo \\
   NGC~4470     & Q12270+0806 &   15.25 & 0.21 &  1.85 &   4.49 &   1.12 &   
Sa? & 34.36 &  13.02 & Virgo \\
\noalign{\smallskip}
\hline
\end{tabular}
$$
$\dagger$ Pair of galaxies not resolved by IRAS.
\end{table}

\newpage
\clearpage

\begin{table}
\caption[]{FIR luminosity and extinction of the four FIRsel galaxies without an 
UV
counterpart.}
\label{threeFIR}
$$
\begin{tabular}{lrr}
\hline
\noalign{\smallskip}
PSCz id. & $A_{0.2}$ & $L_{60}$ \\
         & (mag)      & (erg~s$^{-1}$) \\
\noalign{\smallskip}
\hline
\noalign{\smallskip}
Q02378+3829 & $\geq 4.6$ & $9.0 \times 10^{44}$ \\
Q09031+7855 & $\geq 3.9$ & $1.4 \times 10^{45}$ \\
Q13074+2852 & $\geq 6.0$ & $6.0 \times 10^{43}$ \\
Q12217+0848 & $\geq 3.8$ & $2.3 \times 10^{43}$ \\
\noalign{\smallskip}
\hline
\end{tabular}
$$
\end{table}

\newpage
\clearpage

\setcounter{figure}{0}

   \begin{figure}[t]
   \centering
\caption{$5 \times 5$~arcmin$^{2}$ extracts of the UV (FOCA) and
optical (DSS) blue frames of the UVsel galaxies.}
\label{bonafide}
   \end{figure}





\clearpage

   \begin{figure}[t]
   \centering
\includegraphics[width=16cm]{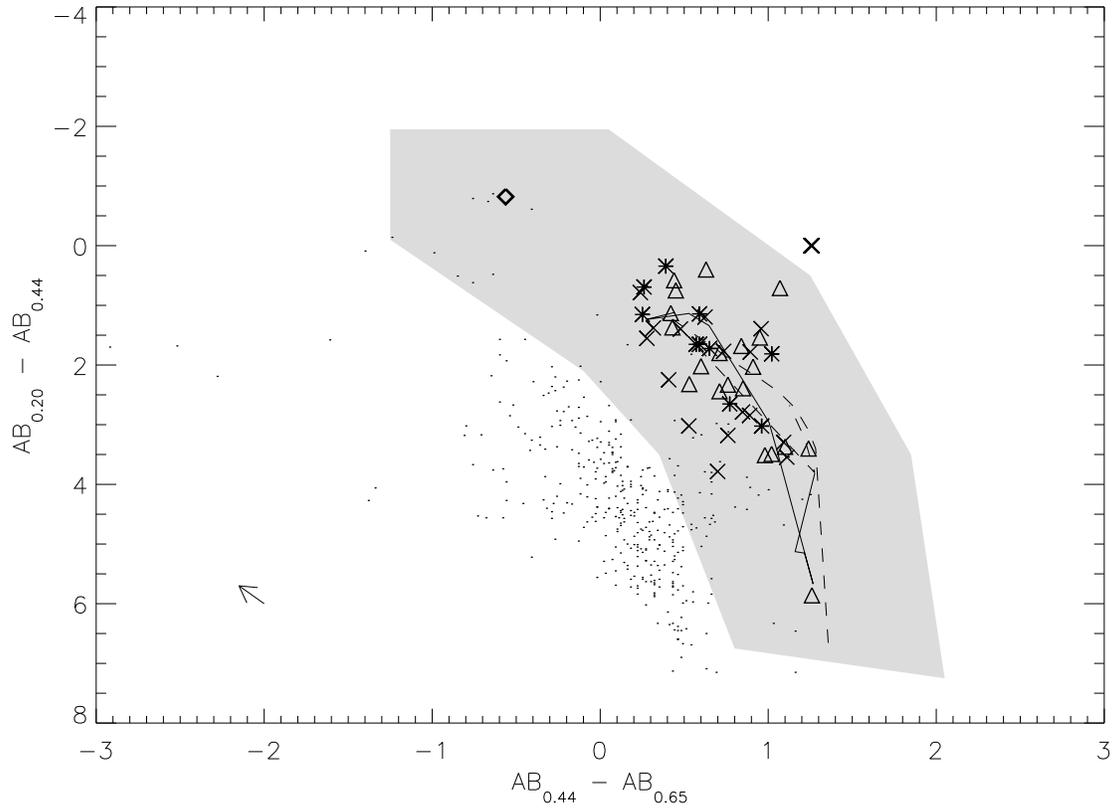}
\caption{$AB_{0.2} - AB_{0.44}$ vs. $AB_{0.44} - AB_{0.65}$ diagram
for the UV sources (points). Asterisks are the starbursts galaxies from Gordon 
et
al. (1997). Triangles are galaxies from the Coma cluster from Donas et
al. (1995). Crosses are the FAUST galaxies from Deharveng et al. (1994).
Solid and dashed lines represent synthetic models of
galaxies following the Hubble sequence from PEGASE (Fioc \& Rocca-Volmerange 1997) and 
Boselli et al. (2003) respectively. 
The bold square represents the position of an instantaneous 
burst of star formation 3~Myr old,
as determined from Starburst99. The shaded region corresponds to the approximate loci
occupied by the real galaxies. The data are not corrected for Galactic
extinction. The arrow at the bottom left corner shows the
extinction vector corresponding to the maximum Galactic extinction for
our UV sources.}
         \label{br}
   \end{figure}

\clearpage

   \begin{figure}[t]
   \centering
\caption{$2 \times 2$~arcmin$^{2}$ extracts of the DSS plates of the
optical counterparts of the UV sources for which the $AB_{0.2} -
AB_{0.44}$ and $AB_{0.44} - AB_{0.65}$ colors are consistent with
those of galaxies. }
         \label{colorgal}
   \end{figure}

\clearpage

   \begin{figure}[t]
   \centering
\includegraphics[width=13cm]{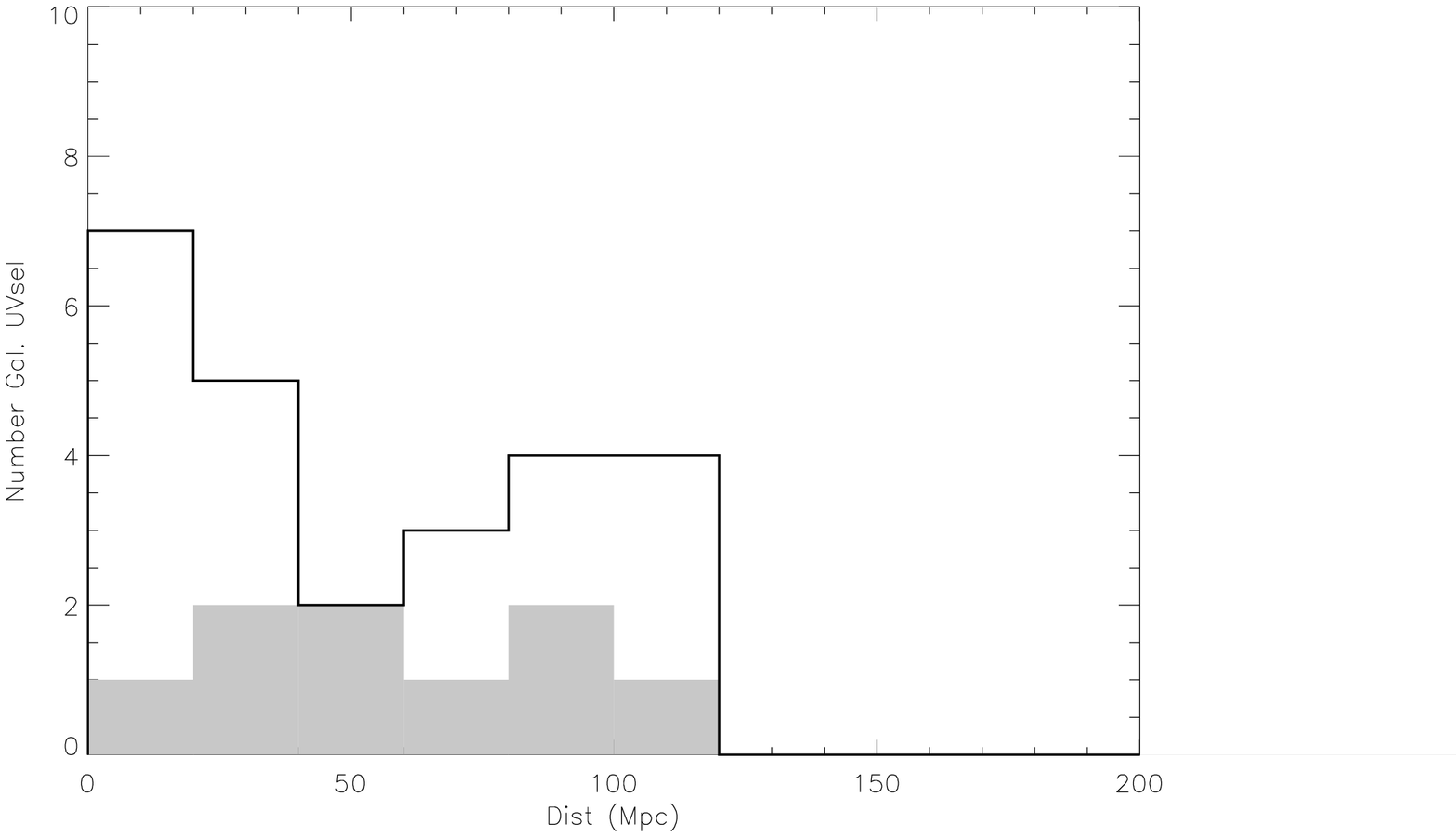}
\caption{Distance distribution of the galaxies in the UVsel sample. The shaded bins correspond to the distribution of the galaxies non associated with clusters.}
\label{distuv}
   \end{figure}

\clearpage

   \begin{figure}[t]
   \centering
\includegraphics[width=13cm]{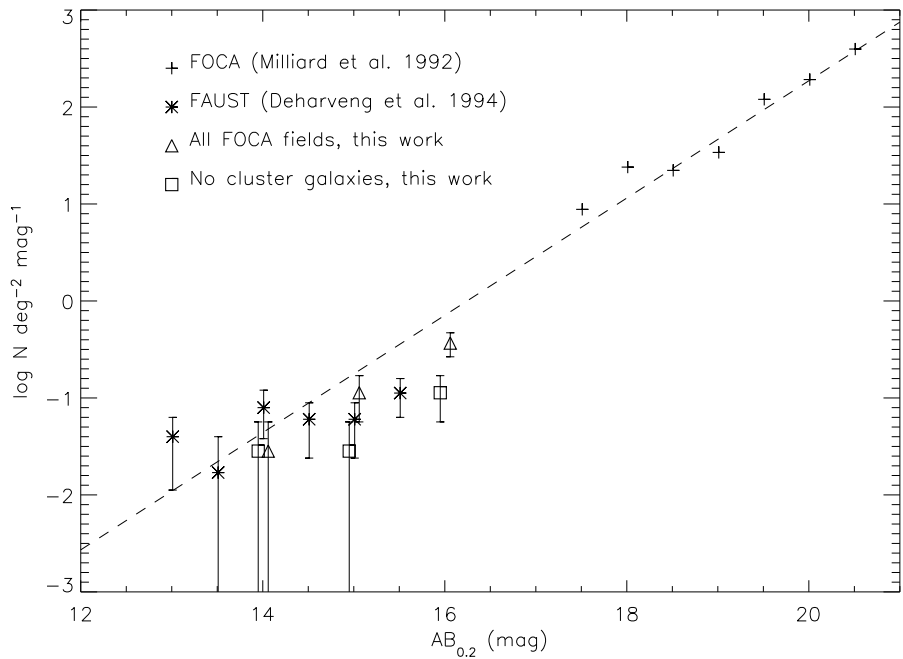}
\caption{Bright UV galaxy counts from several sources of the
literature: Pluses are from Milliard et al. (1992), asterisks from
Deharveng et al. (1994). Open triangles correspond to our total UVsel sample and open squares correspond to the UVsel galaxies not associated to clusters. The symbols corresponding to both sets of
data were slightly shifted along the X-axis in order to avoid superposition.
The dashed straight line corresponds
to the extrapolation to the FOCA counts of Milliard et al. with slope 0.6.}
\label{counts}
   \end{figure}

\clearpage

   \begin{figure}[t]
   \centering
\includegraphics[width=13cm]{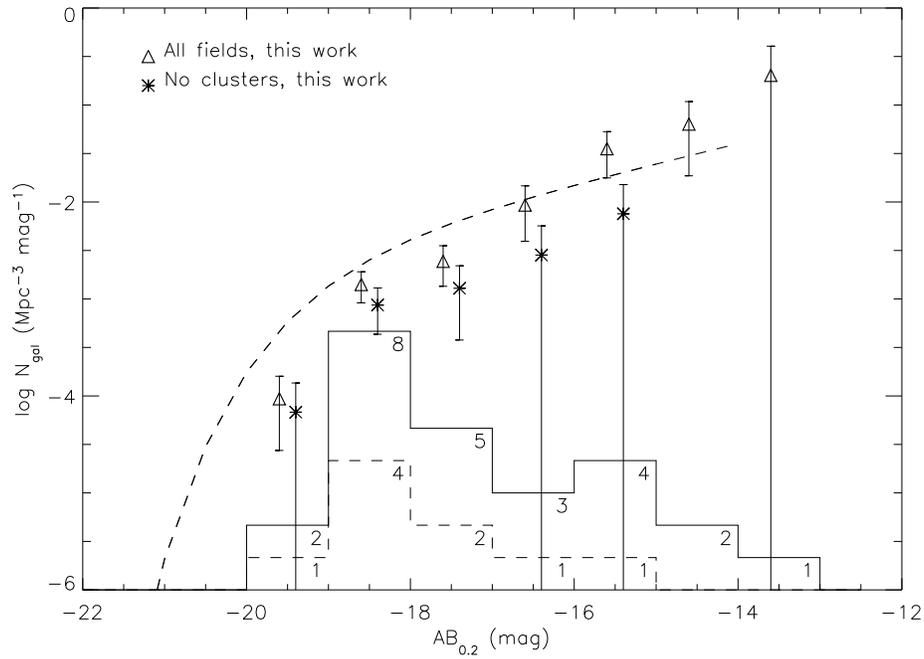}
\caption{UV luminosity function of the total UVsel sample (open triangles) and excluding the cluster galaxies (asterisks). The symbols corresponding to both sets of
data were slightly shifted along the X-axis in order to avoid superposition.
The dashed line corresponds to
the best fit to a Schechter function of the luminosity function of
Sullivan et al. using the $1/V_{max}$ method. 
UV magnitudes are not corrected for internal
extinction. The histograms show the magnitude distributions of the total sample (solid line) and the subsample without cluster galaxies (dashed line). The number of galaxies in each bin are indicated at the top of the bins.}
\label{uvlf}
   \end{figure}

\clearpage

   \begin{figure}[t]
   \centering
\includegraphics[width=13cm]{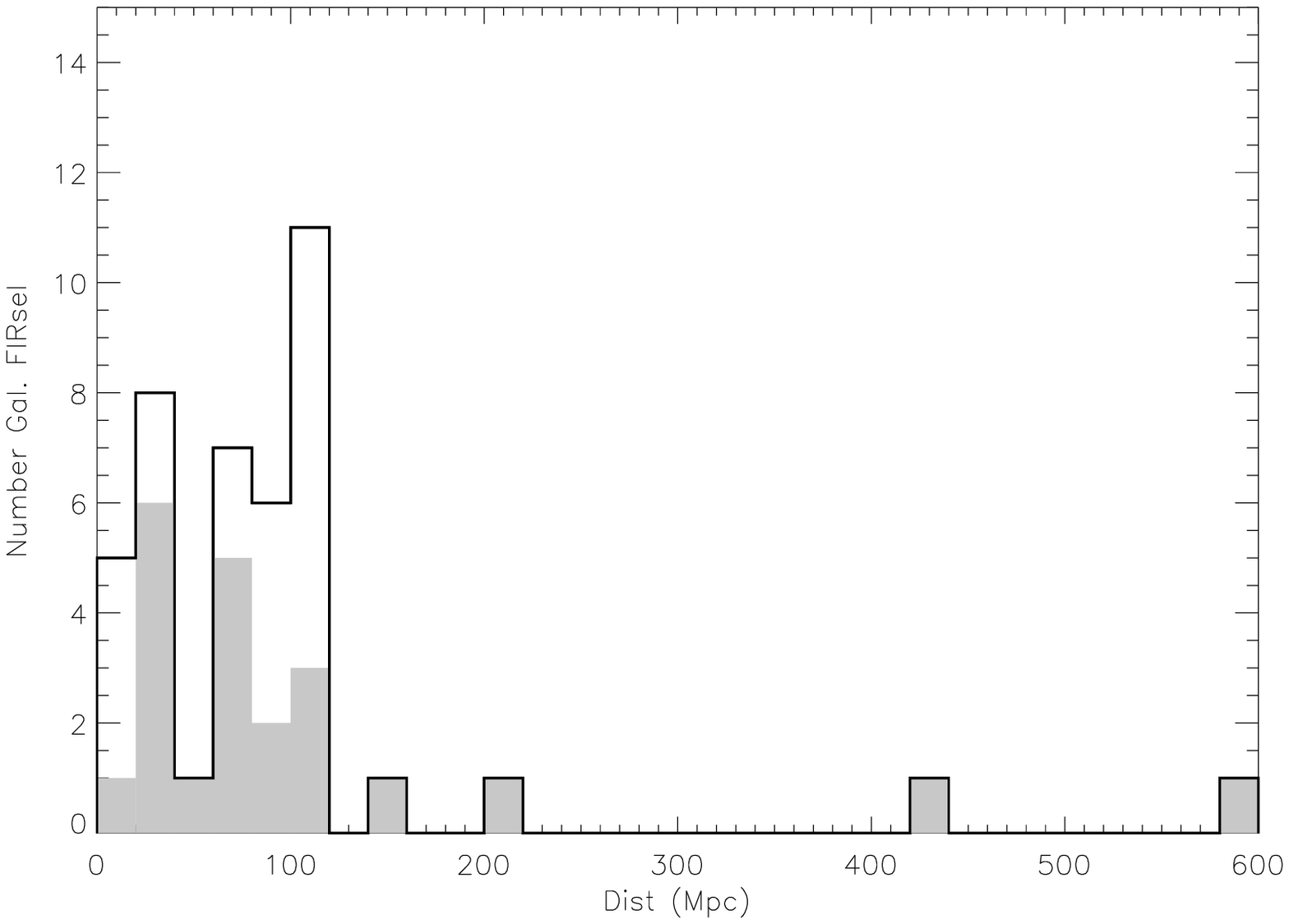}
\caption{Distance distribution of the galaxies in the FIRsel sample. The shaded bins correspond to the distribution of the galaxies non associated with clusters.}
\label{distfir}
   \end{figure}

\clearpage

   \begin{figure}[t]
   \centering
\includegraphics[width=13cm]{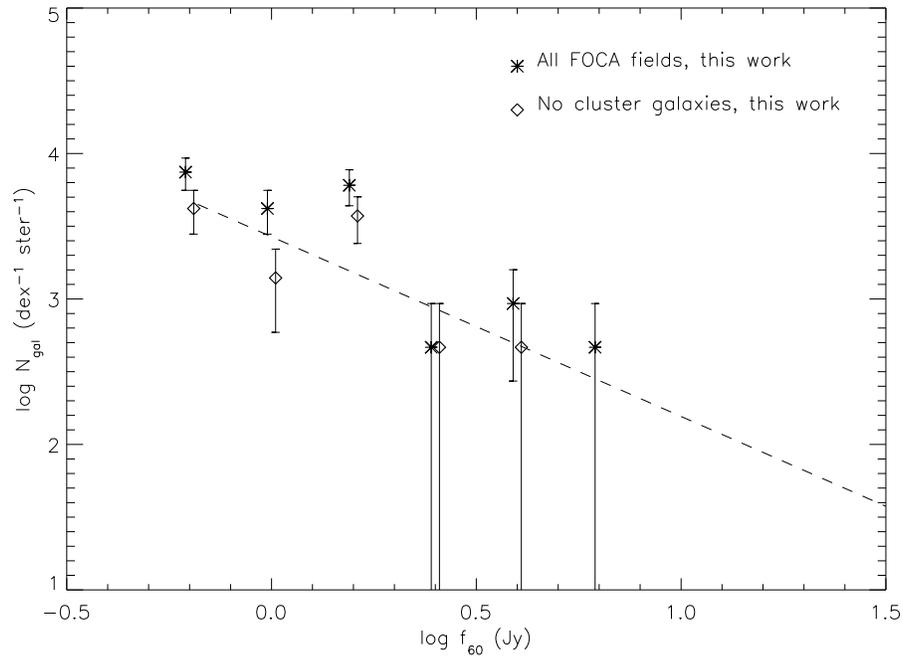}
\caption{Galaxy counts as a function of $f_{60}$ for the FIRsel
sample. Asterisks correspond to the total FIRsel sample whereas open diamonds correspond to the galaxies of the FIRsel sample not associated to clusters. The symbols corresponding to both sets of
data were slightly shifted along the X-axis in order to avoid superposition.
The dashed line corresponds to average counts presented by
Saunders et al. (2000) for three different subsamples of the PSCz.}
\label{fircounts}
   \end{figure}

\clearpage

   \begin{figure}[t]
   \centering
\includegraphics[width=13cm]{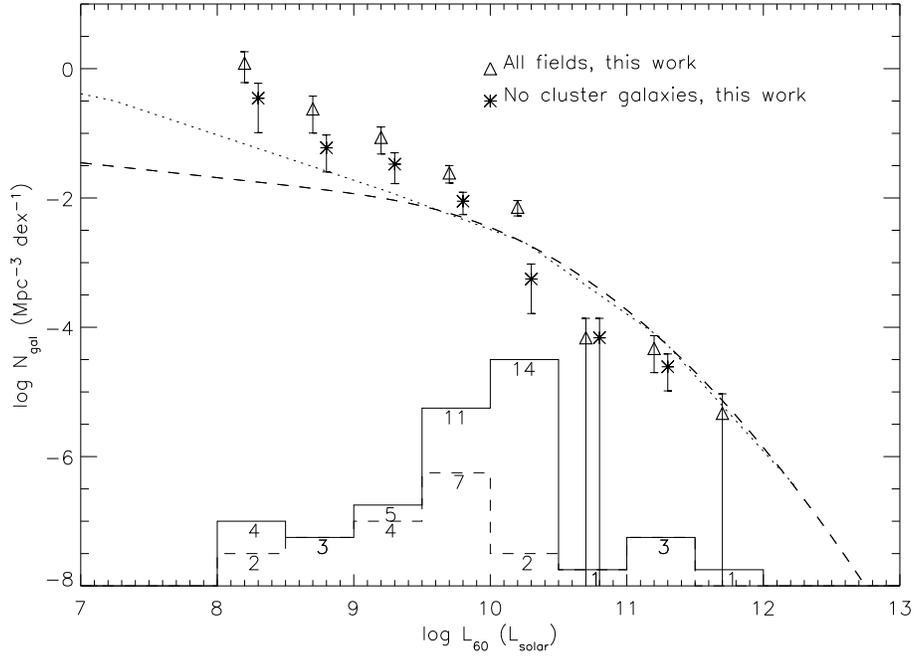}
\caption{Luminosity functions of the total FIRsel sample (open triangles) and of the subsample without cluster galaxies (asterisks) as a function of the
$L_{60}$ luminosity. The symbols corresponding to both sets of
data were slightly shifted along the X-axis in order to avoid superposition.
The dashed and dotted line correspond to the LF of Takeuchi et al. (2003) computed using an analytic method and the $1/V_{max}$ method respectively.
The histograms show the luminosity distributions of the total sample (solid line) and the subsample without cluster galaxies (dashed line). The number of galaxies in each bin are indicated at the top of the bins.}
\label{firlf}
   \end{figure}

\clearpage

   \begin{figure}[t]
   \centering
\includegraphics[width=14cm]{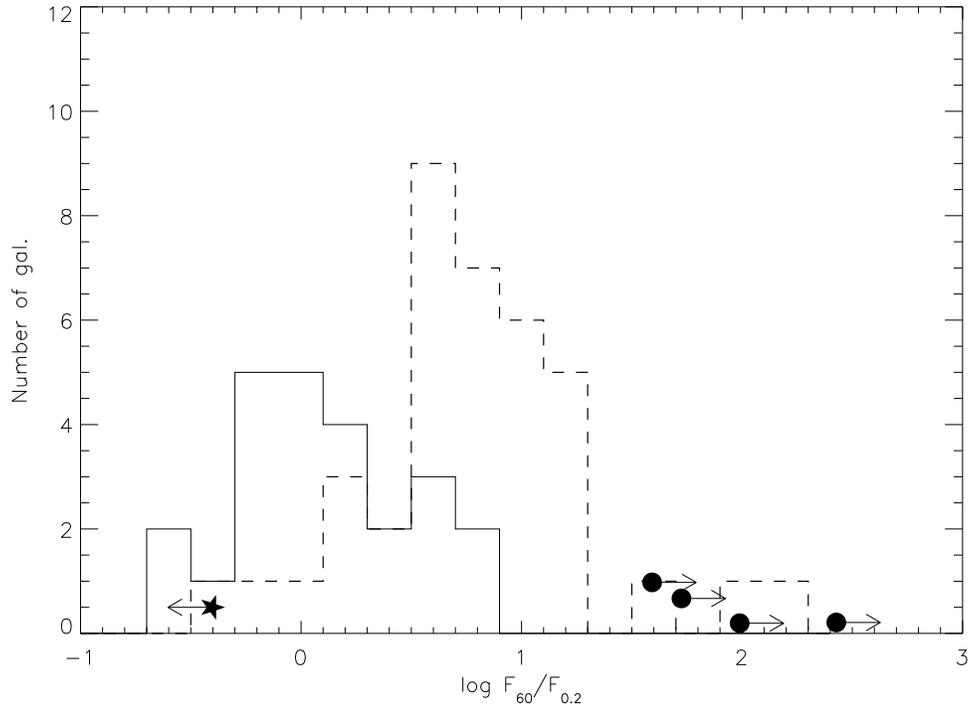}
\caption{Histograms of $\log(F_{60}/F_{0.2})$ for the UVsel sample
(dashed line) and the FIRsel sample (solid line). The galaxy of the UVsel galaxy with no FIR counterpart is indicated by a star as an upper limit. The four galaxies
from the FIRsel sample with no UV counterpart are indicated with
filled dots as lower limits.}
         \label{histfir_fuv}
   \end{figure}

\clearpage

   \begin{figure}[t]
   \centering
\includegraphics[width=14cm]{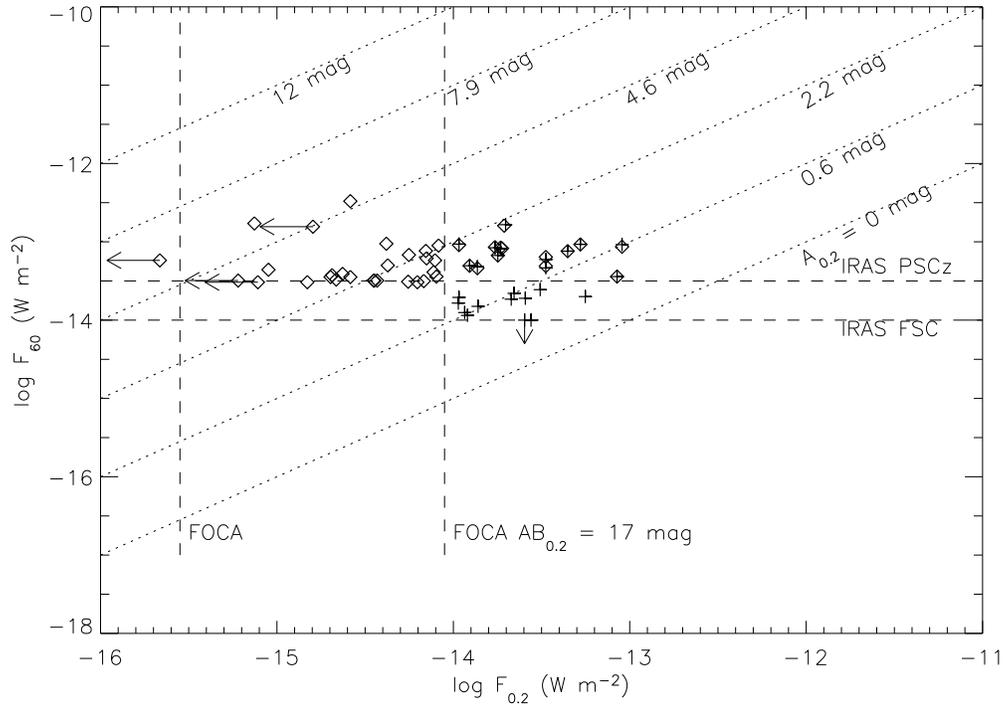}
\caption{$\log F_{60}$ vs. $\log F_{0.2}$ for the FIRsel sample
(open diamonds) and the UVsel sample (pluses). The galaxy from the UVsel sample without a FIR counterpart as well as the four
galaxies from the FIRsel sample without a detection in UV are
represented as upper limits with arrows. Dashed horizontal lines
indicate the limiting fluxes of the FOCA experiment, and of the UVsel
sample. Dashed vertical lines indicate the limiting fluxes of the FSC
and PSCz catalogs. Lines indicating different values of the extinction
at 0.2$\mu$m are also represented by dotted lines.}
         \label{plotfir_fuv}
   \end{figure}

\clearpage

   \begin{figure}[t]
   \centering
\includegraphics[width=16cm]{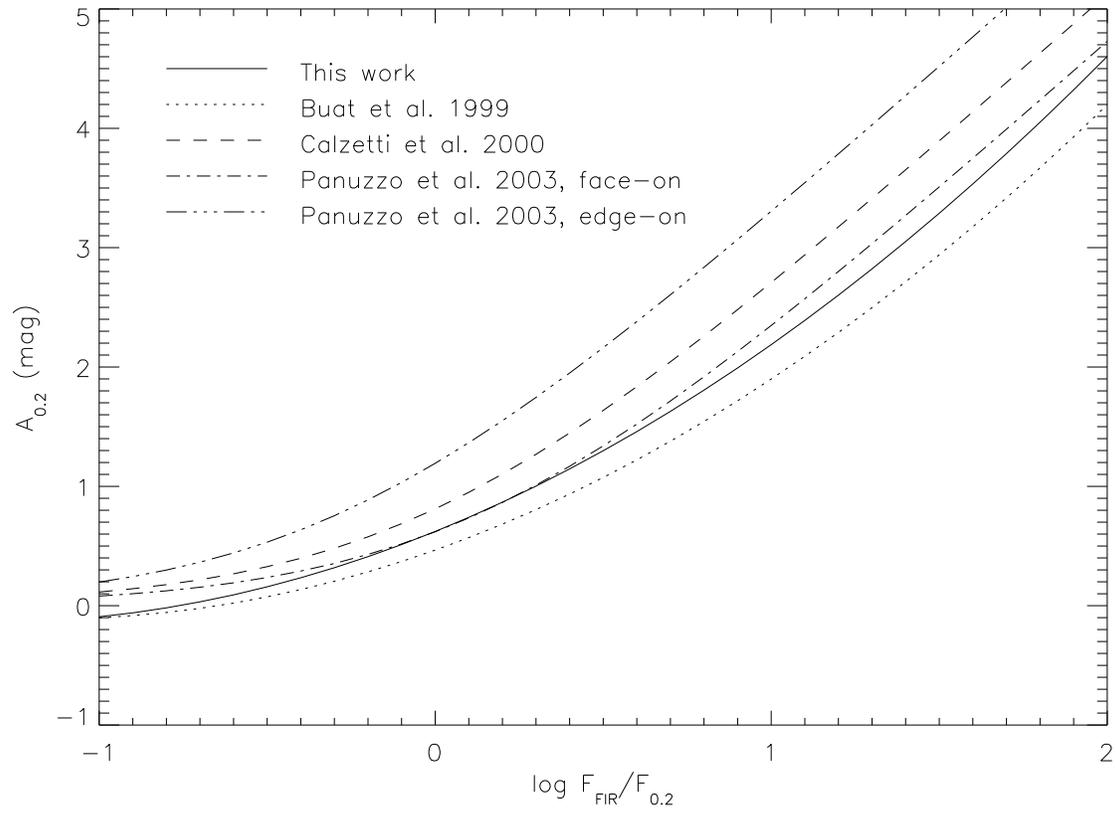}
\caption{Comparison of different extinction corrections as a function
of the $F_{\mbox{\scriptsize FIR}}/F_{\mbox{\scriptsize UV}}$
ratio. The solid line corresponds to the one used throughout this
work.}
         \label{compa_extin}
   \end{figure}

\clearpage

   \begin{figure}[t]
   \centering
\includegraphics[width=8.95cm]{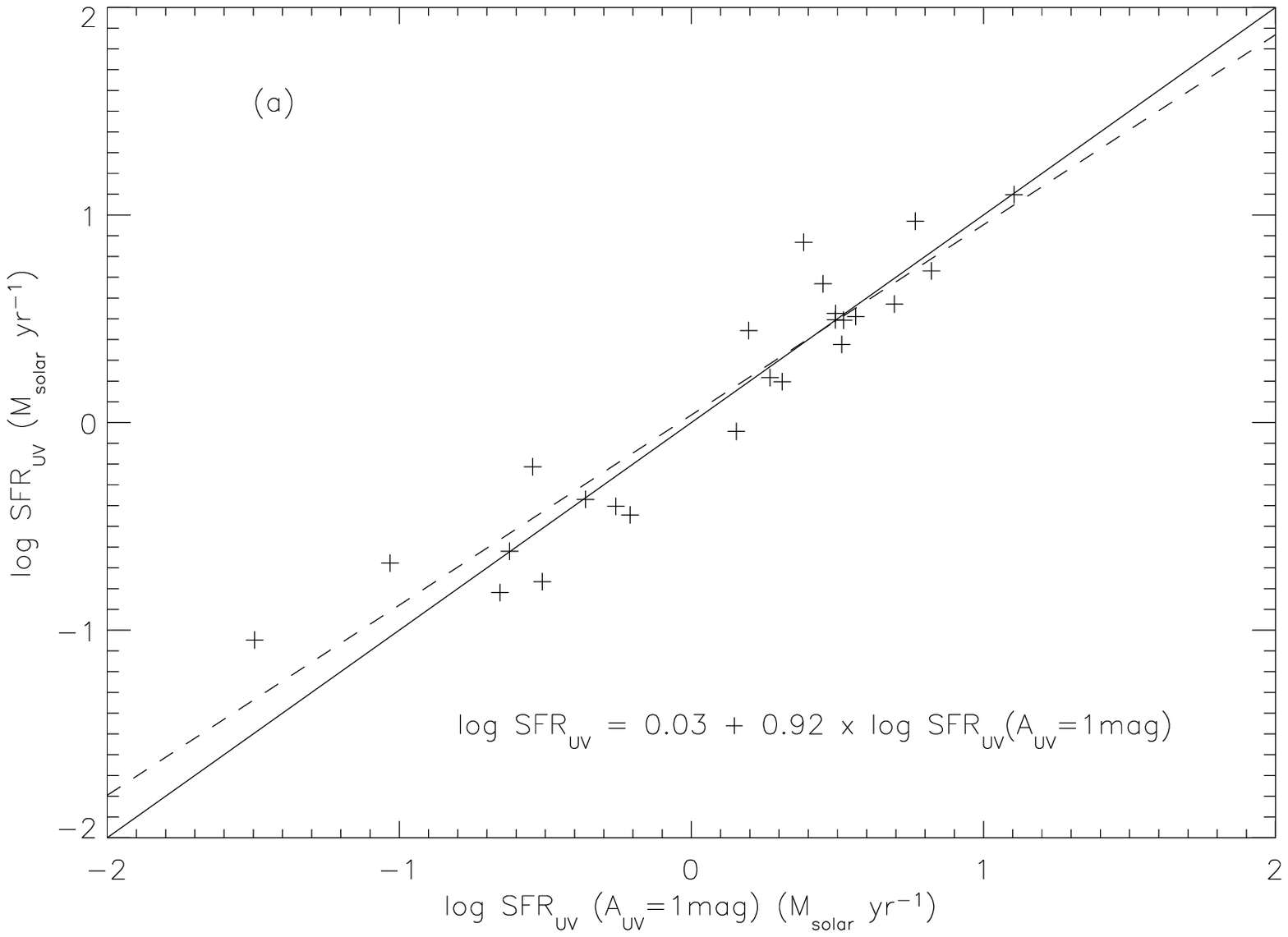}
\includegraphics[width=8.95cm]{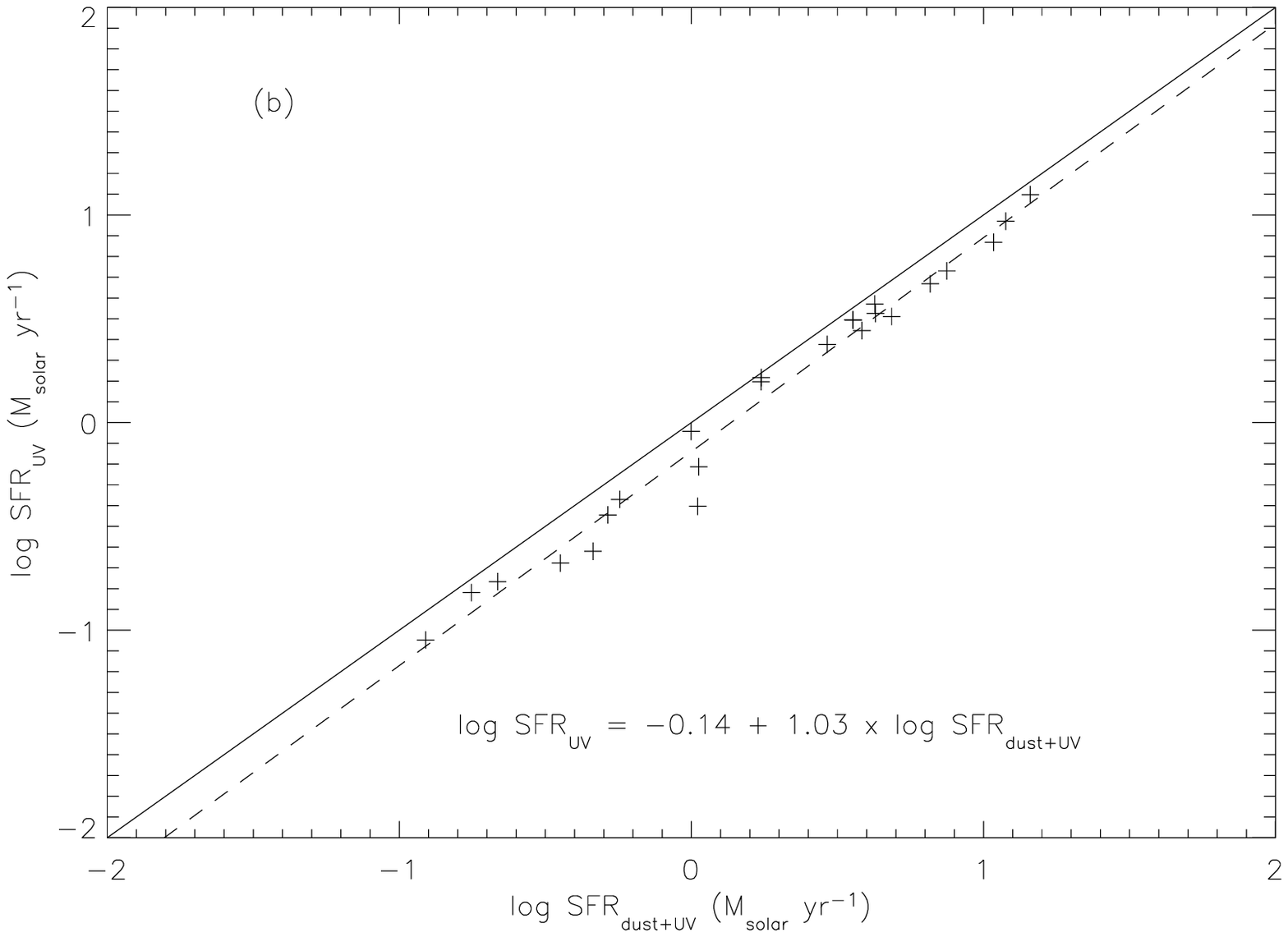}
\includegraphics[width=8.95cm]{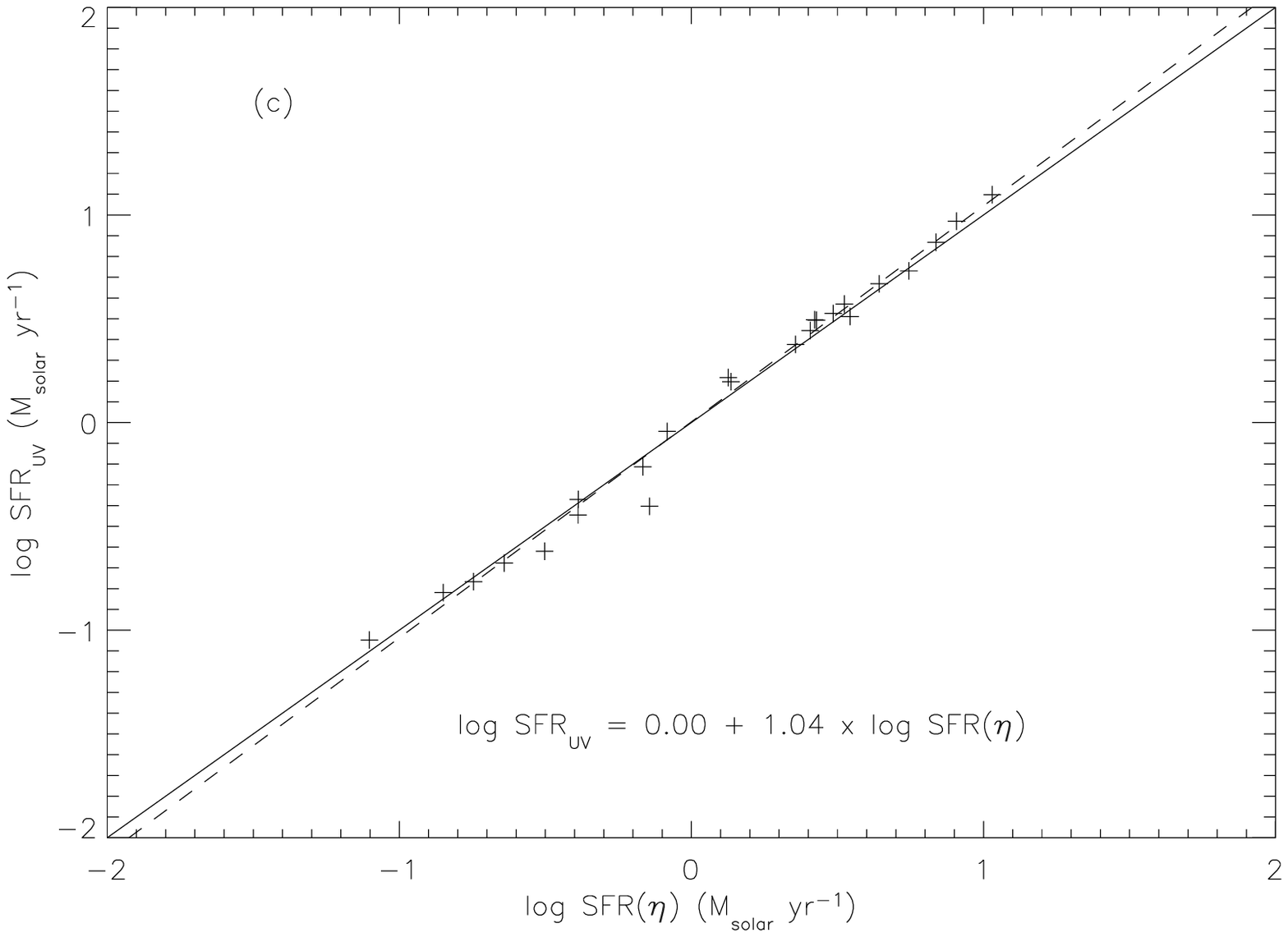}
\includegraphics[width=8.95cm]{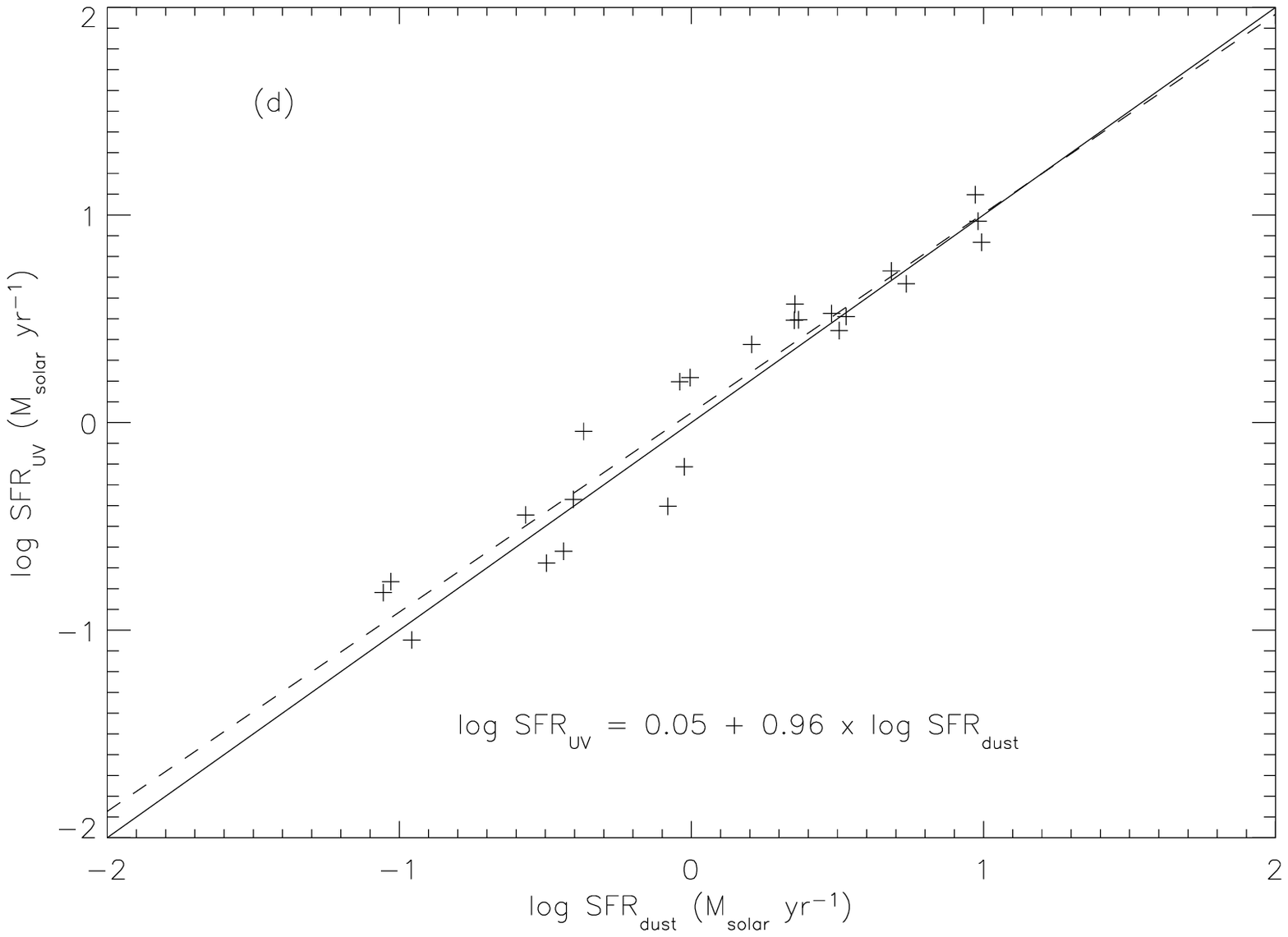}
\caption{Comparison of the different estimations of the SFR for the galaxies of 
the
UVsel sample: (a) $\log SFR_{\mbox{\scriptsize 
UV}}(A_{0.2}=1.07)/SFR_{\mbox{\scriptsize UV}}$
vs. $\log SFR_{\mbox{\scriptsize UV}}$, (b) $\log SFR_{\mbox{\scriptsize dust$+$UV}}$ 
vs. $\log SFR_{\mbox{\scriptsize UV}}$, (c) $\log SFR(\eta)$ vs. $\log SFR_{\mbox{\scriptsize UV}}$, (d) $\log 
SFR_{\mbox{\scriptsize dust}}$
vs. $\log SFR_{\mbox{\scriptsize UV}}$. SFRs are always expressed
in $M_{\odot}$~yr$^{-1}$. Dashed lines represent the best
fits to straight lines. The galaxy without a FIR counterpart is excluded.
}
         \label{sfr_UVsel}
   \end{figure}

\clearpage

   \begin{figure}[t]
   \centering
\includegraphics[width=8.95cm]{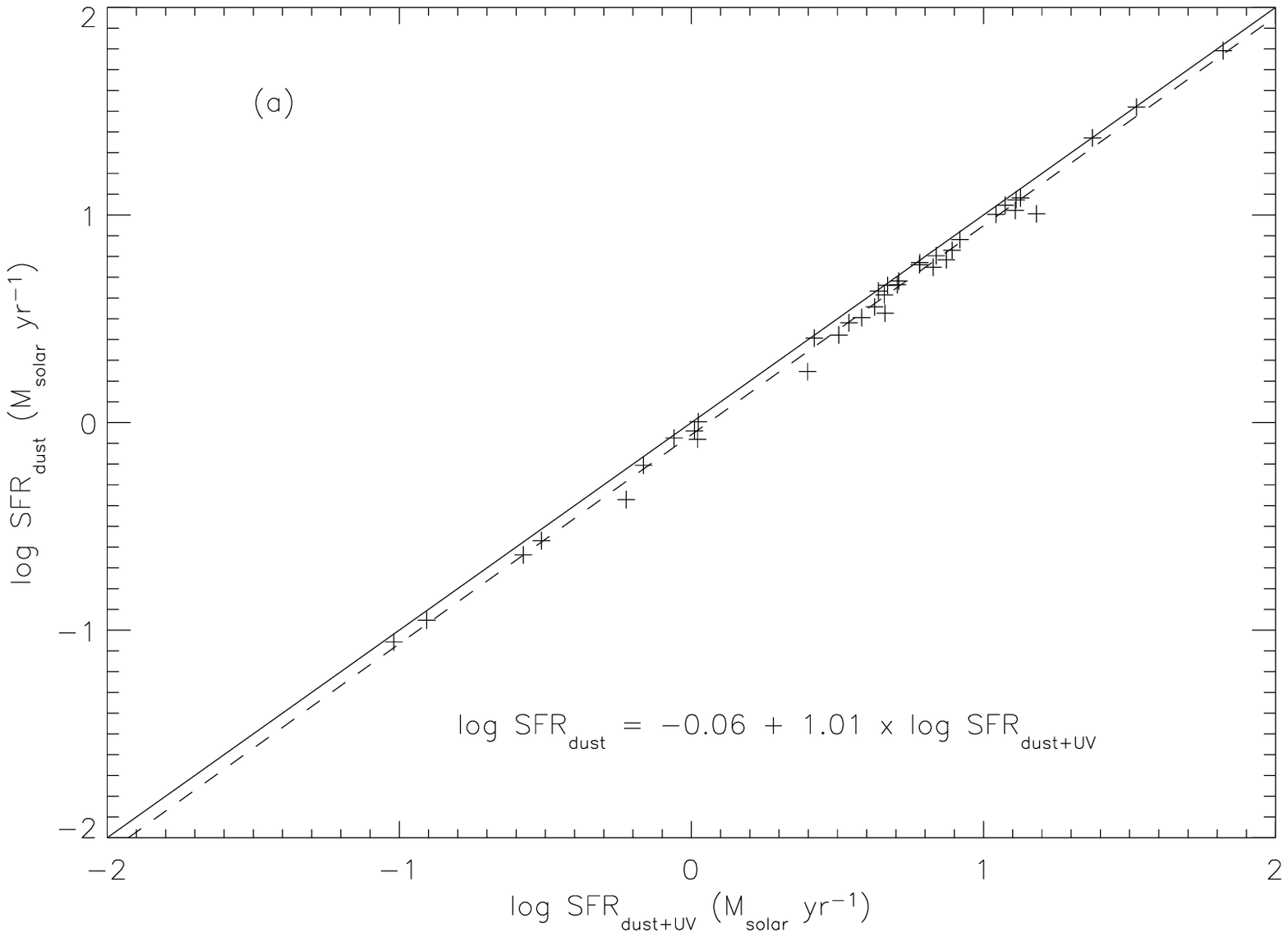}
\includegraphics[width=8.95cm]{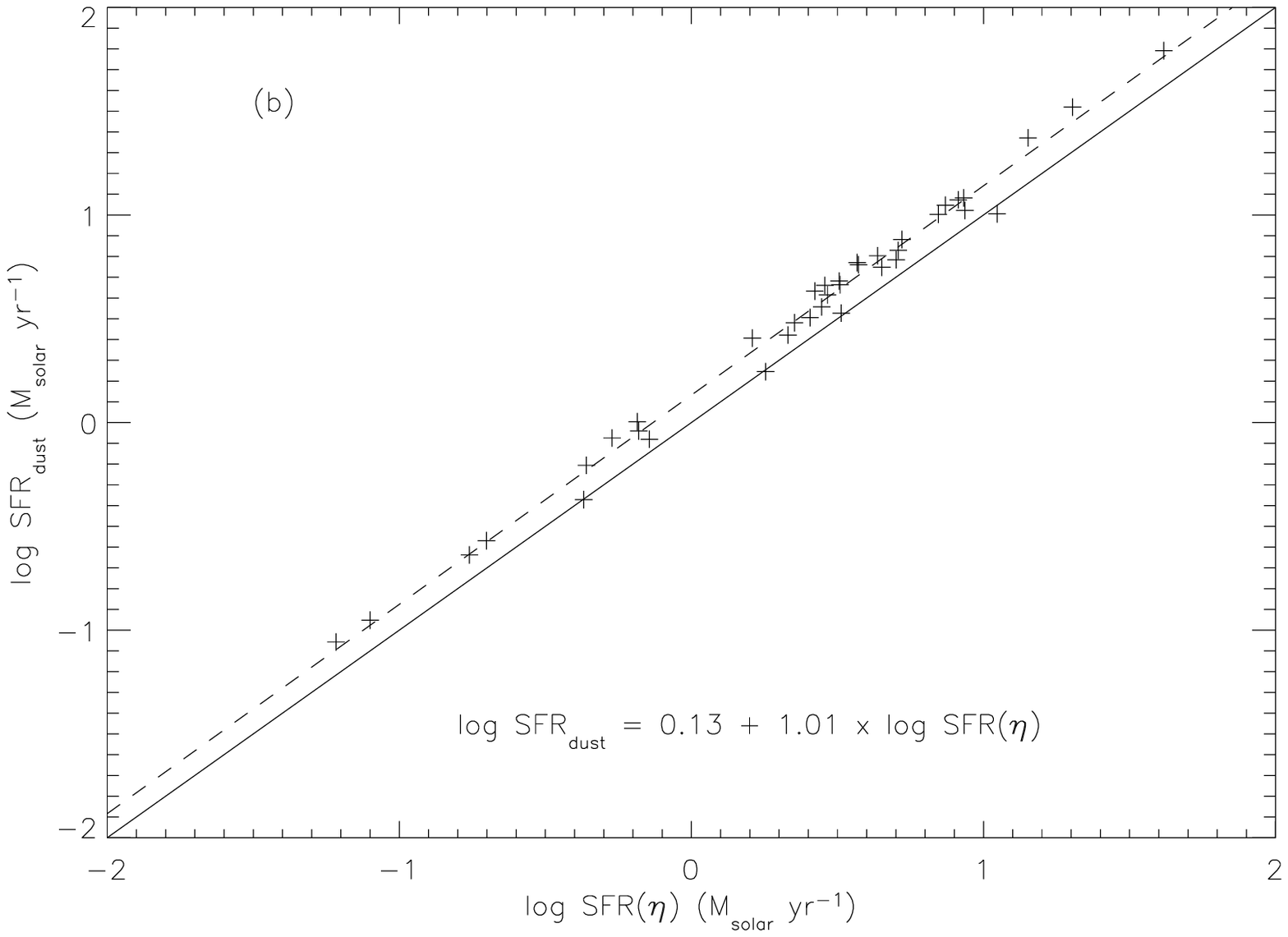}
\includegraphics[width=8.95cm]{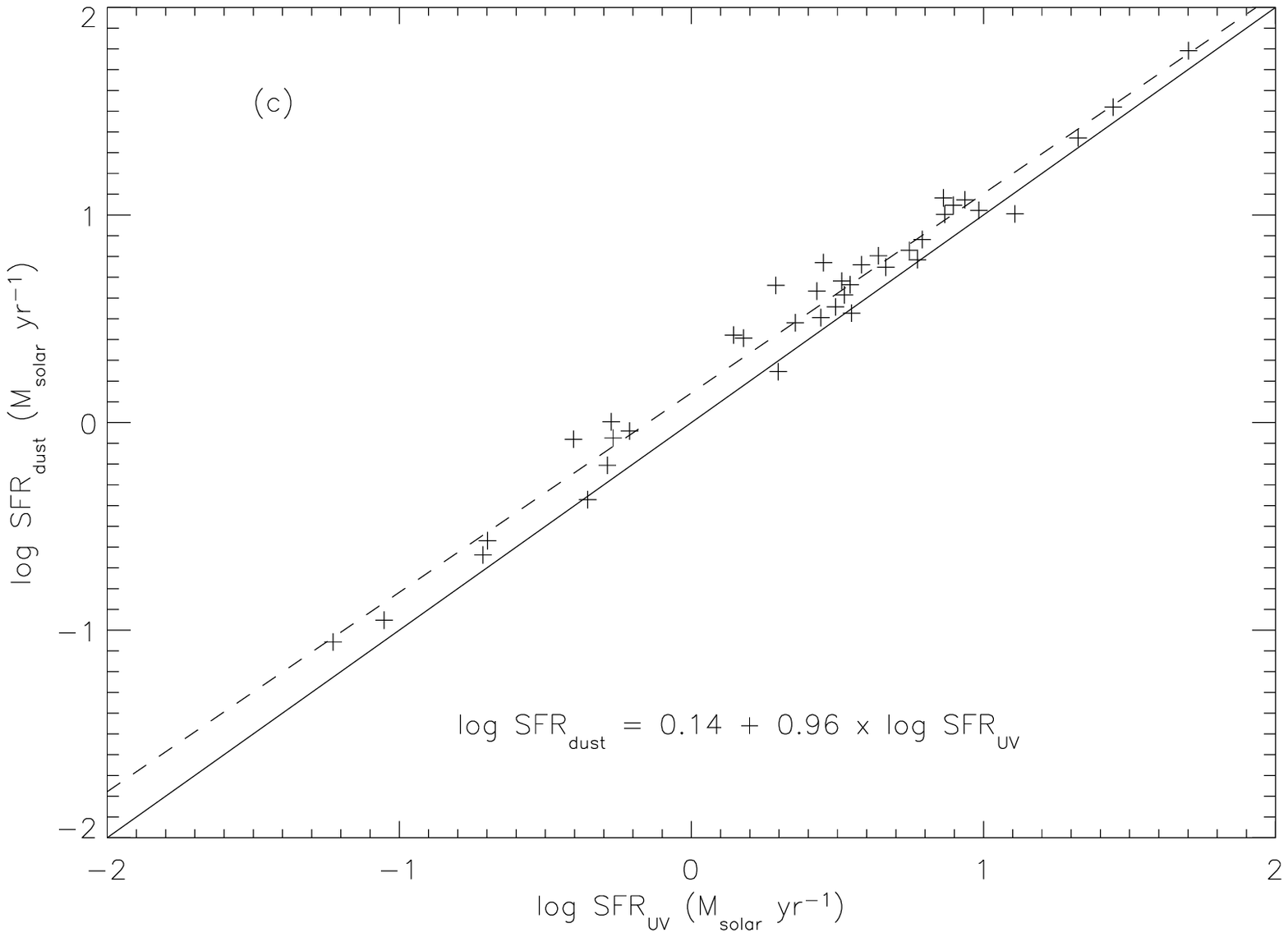}
\includegraphics[width=8.95cm]{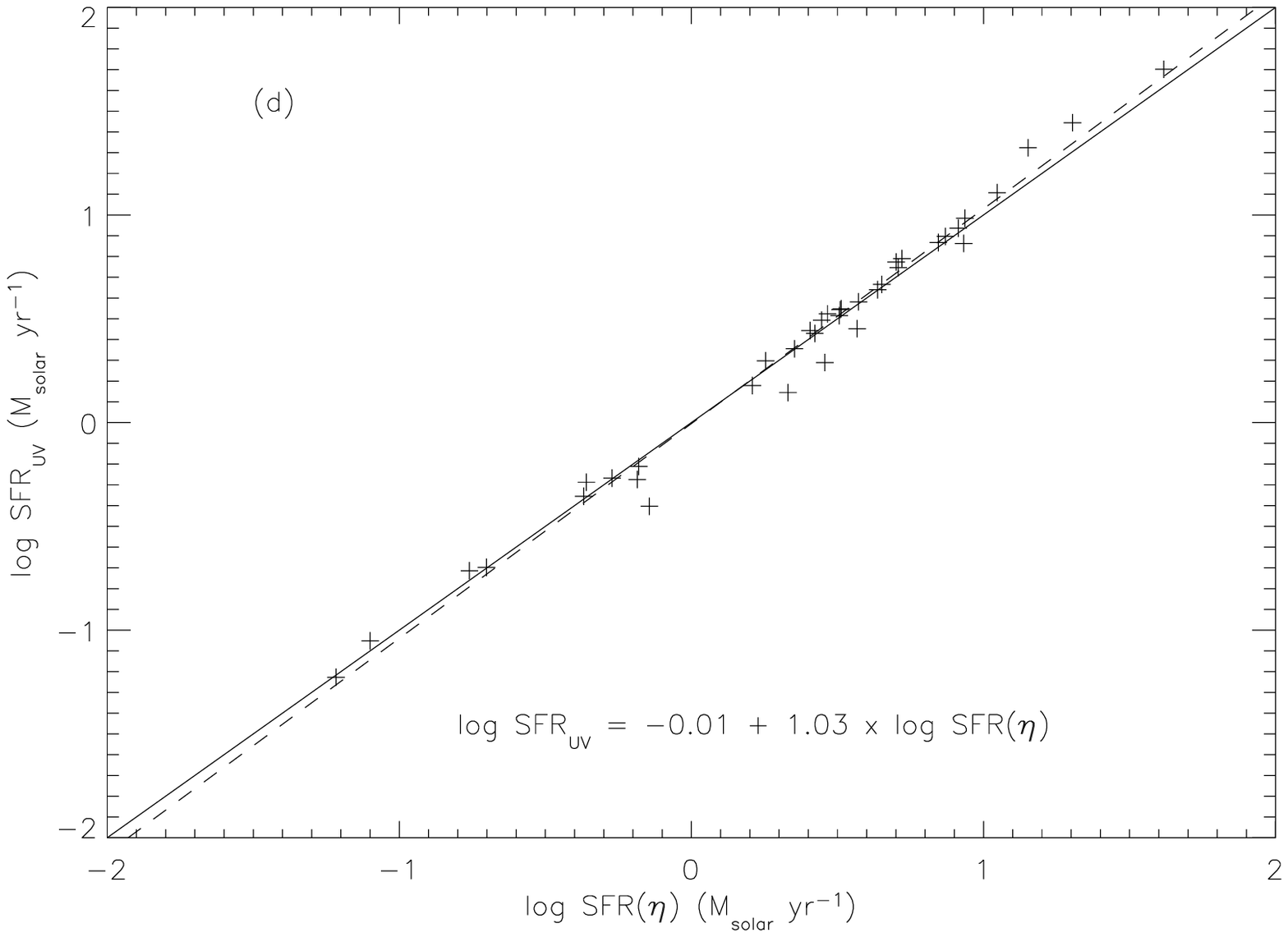}
\caption{Comparison of the different estimations of the SFR for the galaxies of 
the
FIRsel sample: (a) $\log SFR_{\mbox{\scriptsize dust$+$UV}}$ vs. $\log
SFR_{\mbox{\scriptsize dust}}$, (b) $\log SFR(\eta)$ vs. 
$\log
SFR_{\mbox{\scriptsize dust}}$, (c) $\log SFR_{\mbox{\scriptsize UV}}$ vs. $\log 
SFR_{\mbox{\scriptsize dust}}$, (d) $\log
SFR(\eta)$ vs. $\log SFR_{\mbox{\scriptsize
UV}}$. SFRs are always expressed in $M_{\odot}$~yr$^{-1}$. Dashed
lines represent the best fits to straight lines. The four galaxies without an UV counterpart are excluded.}
         \label{sfr_firsel}
   \end{figure}

\clearpage

   \begin{figure}[t]
   \centering
\includegraphics[width=14cm]{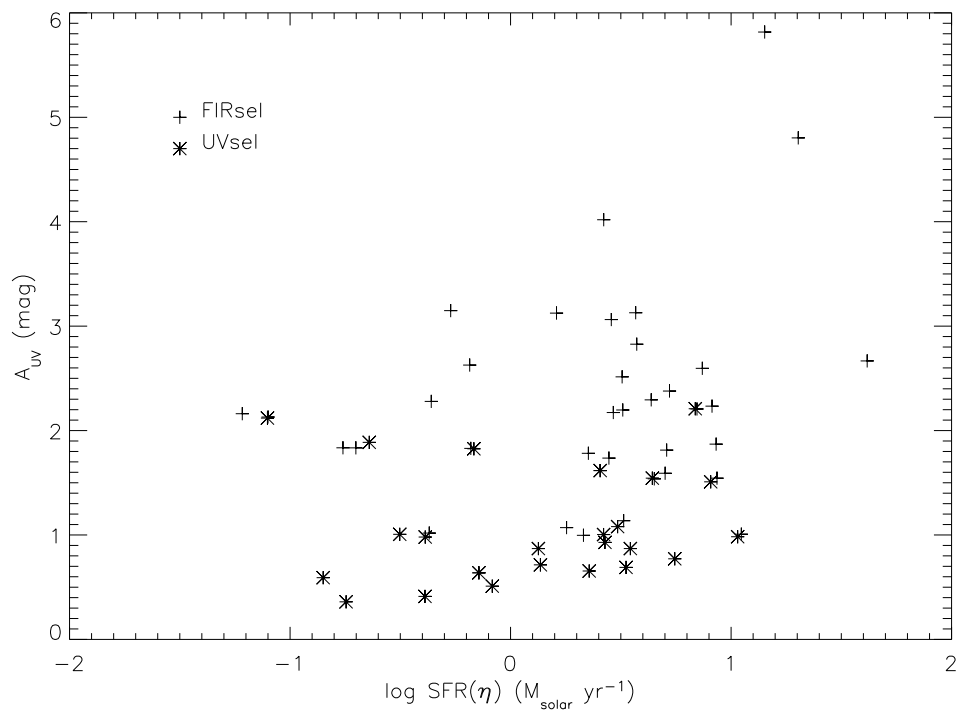}
\caption{$\log SFR(\eta)$ vs. $A_{\mbox{\scriptsize
UV}}$ for the UVsel (asterisks) and FIRsel (pluses) samples. SFRs are
expressed in $M_{\odot}$~yr$^{-1}$. The UVsel galaxy without a FIR counterpart as well as the four FIRsel galaxies without an UV counterpart are excluded.}
         \label{sfrtot_fdustfuv}
   \end{figure}

\clearpage

   \begin{figure}[t]
   \centering
\includegraphics[width=14cm]{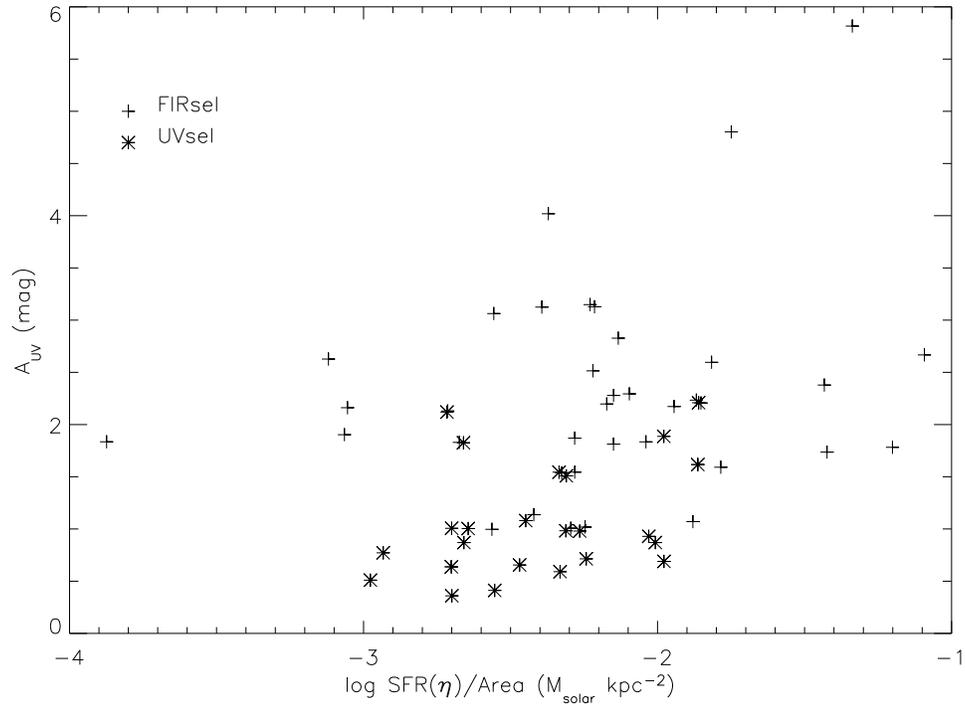}
\caption{Same as Fig~\ref{sfrtot_fdustfuv} with the
$SFR(\eta)$ normalized to the optical area of the
galaxies in kpc$^{2}$. Symbols are also as in
Fig~\ref{sfrtot_fdustfuv}.}
         \label{sfrtot_s_fdustfuv}
   \end{figure}

\newpage

\clearpage

\appendix

\section{Photometric quality of the USNO-B1.0 catalog}

\begin{figure}[!hb]
   \centering
\includegraphics[width=12cm]{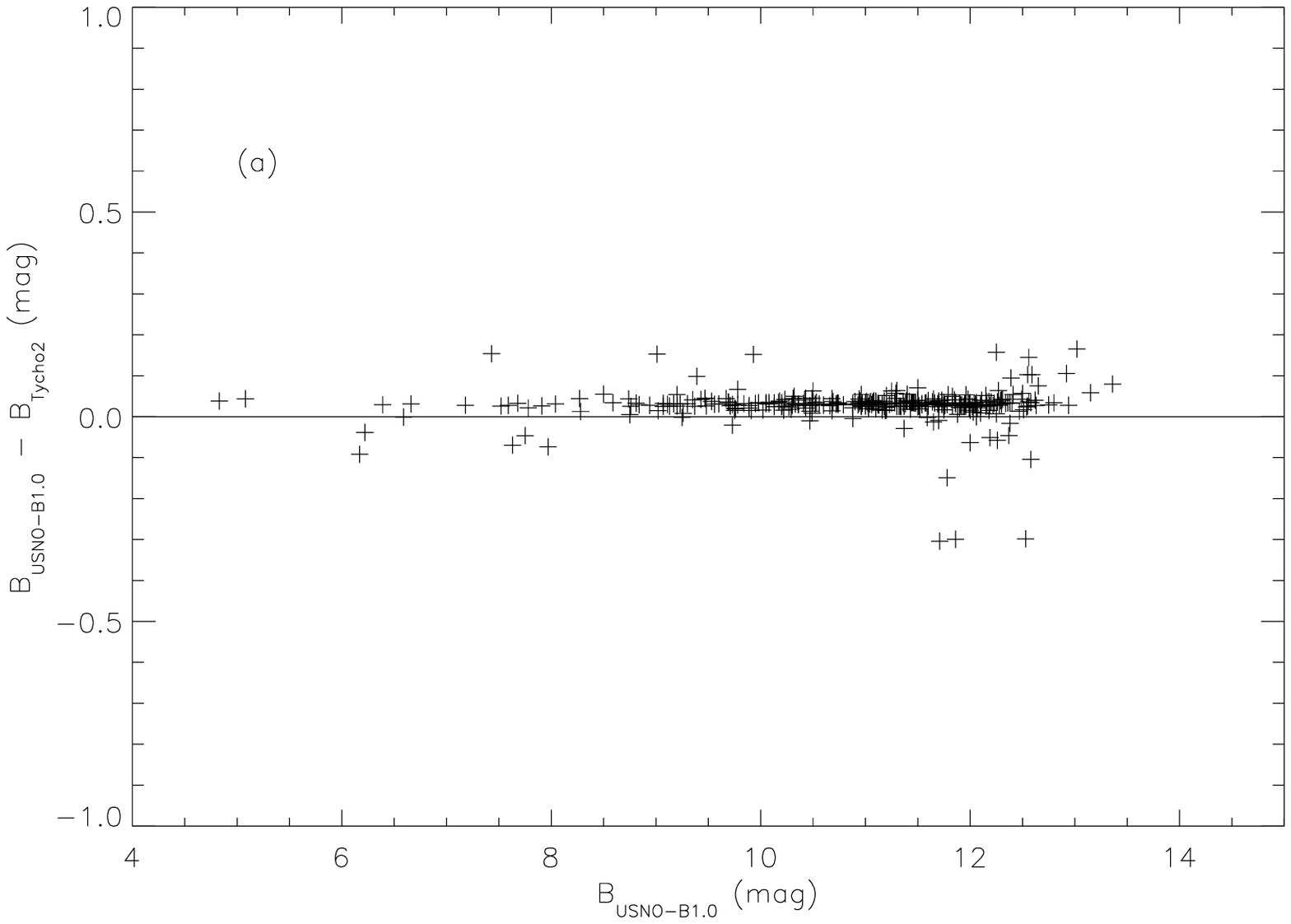}
\includegraphics[width=12cm]{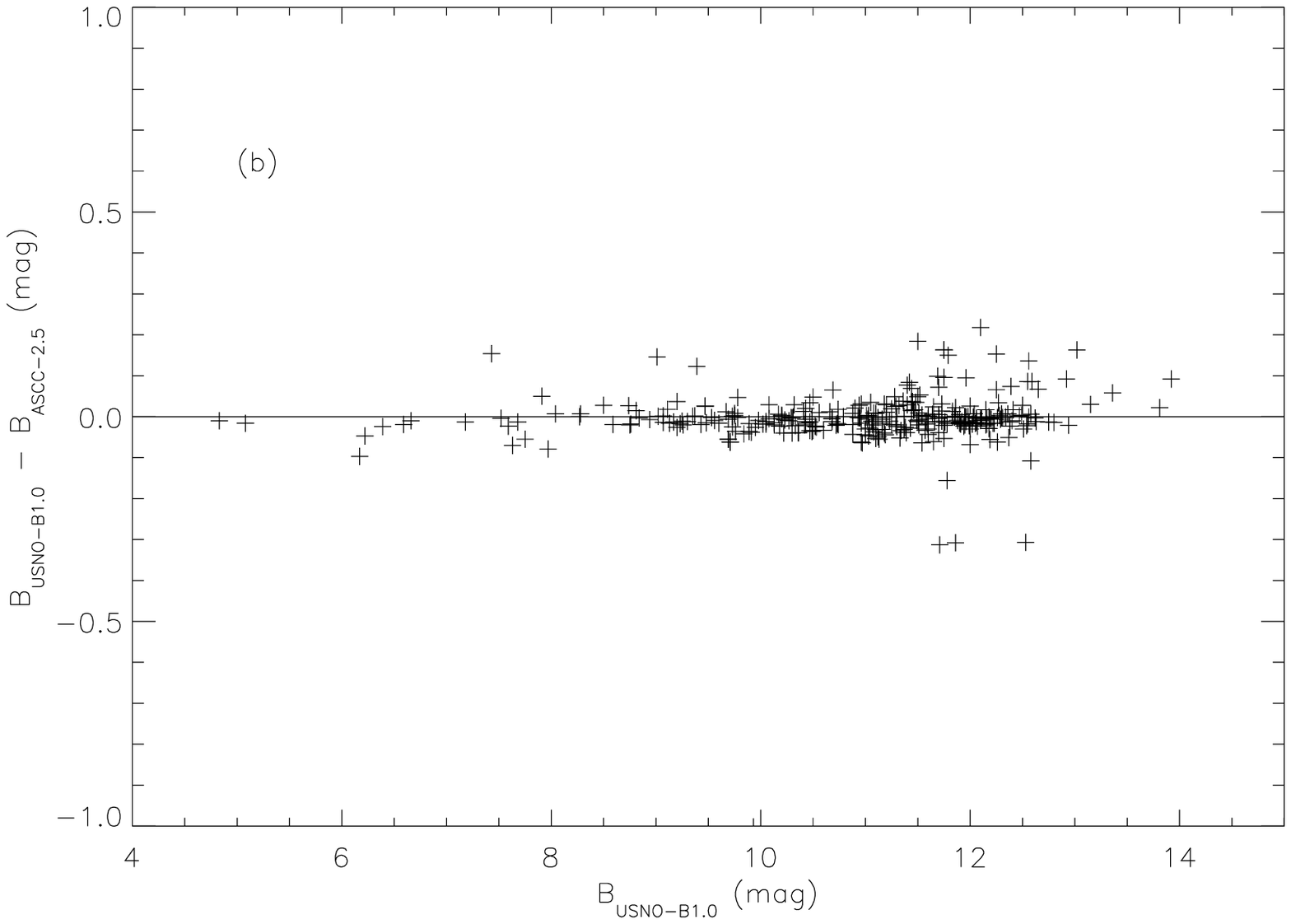}
\caption{Comparison between the $B$ photometry of the USNO-B1.0 catalog with
other catalogs: Tycho2 (a) and ASCC-2.5 (b).}
         \label{compa}
   \end{figure}

The USNO-B1.0 catalog is a compilation from the digitization of
various photographic sky surveys plates by the Precision Measuring
Machine (PMM) located at the US Naval Observatory Flagstaff Station
(NOFS). Details about the data handling and photometric calibration
can be found in Monet et al. (2003).
We show in Fig~\ref{compa} the comparison of the $B$-band
photometry of the USNO-B1.0 objects in common with Tycho-2 (H{\o}g et
al. 2000) and ASCC-2.5 (Kharchenko 2001). As shown by this figure, the
agreement between the photometry of the three catalogs is quite good
(within 0.3~mag). This is not surprising since the Tycho-2 catalog
was used to calibrate the brightest objects of USNO-B1.0 and since
ASCC-2.5 partially overlaps Tycho-2. But even for the ASCC-2.5 objects
not present in Tycho-2, we show that the agreement between the
photometry of ASCC-2.5 and USNO-B1.0 is good enough so that we can
rely on the USNO-B1.0 photometry.
Concerning the $R$-band photometry of the USNO-B1.0 catalog, no
comparison was made because of the lack of available catalogs with
$R$-band photometry. However since the $B$-band photometry is
fairly good, we feel confident about the quality of the $R$-band one.

\clearpage

\clearpage
\newpage

\end{document}